\documentclass[preprint,12pt]{elsarticle}
 \usepackage{amsmath,amssymb}
 \usepackage{epsfig,graphicx}
 
\newcommand{\myfrac}[2]{\frac{ \mbox{$#1$} }{ \mbox{$#2$} }}

\journal{Physical Letters B}
 
\begin{document}
 
\begin{frontmatter}

\title{\bf Search for Heavy Neutrino in $K^-\to\mu^-\nu_h (\nu_h \to \nu \gamma)$ Decay at ISTRA+ Setup}
\author{V.~A.~Duk \fnref{fn1} }
\author{V.~N.~Bolotov,  A.~A.~Khudyakov, V.~A.~Lebedev, \\  A.~I.~Makarov,  V.~P.~Novikov, A.~Yu.~Polyarush}

\address{Institute for Nuclear Research of RAS, Moscow, Russia}

\author{S.~A.~Akimenko, G.~I.~Britvich, A.~P.~Filin, A.~V.~Inyakin,\\
S.~A.~Kholodenko, V.~M.~Leontiev, V.~F.~Obraztsov, V.~A.~Polyakov,\\
V.~I.~Romanovsky, O.~V.~Stenyakin, O.~G.~Tchikilev, V.~A.~Uvarov, O.~P.~Yushchenko}
\address{Institute for High Energy Physics, Protvino, Russia}

\fntext[fn1]{Viacheslav.Duk@cern.ch}


\begin{abstract}

Heavy neutrino $\nu_h$ with $m_h \lesssim$ 300MeV/c$^2$ can be effectively searched for in kaon decays. 
We put upper limits on a mixing matrix element $|U_{\mu h}|^2$ for radiatively decaying $\nu_h$
from $K^-\to\mu^-\nu_h (\nu_h \to \nu \gamma)$ decay chain
in the following parameter region: 30MeV/c$^2 \le m_h \le$ 80MeV/c$^2; 10^{-11}s \le \tau_h \le 10^{-9}$s.
For the whole region  $|U_{\mu h}|^2 \lesssim 5 \cdot 10^{-5}$  for Majorana type of $\nu_h$
and $|U_{\mu h}|^2 \lesssim 8 \cdot 10^{-5}$ for the Dirac case.

\end{abstract}


\end{frontmatter}

\section{Introduction}
\label{sec_one}

For more than ten years latest results of short-baseline neutrino experiments have been widely discussed
and still there is no clear understanding of 
an event excess observed by LSND \cite{lsnd} and MiniBooNE \cite{miniboone1,miniboone2} experiments and their 
contradiction with KARMEN \cite{karmen} results.

Oscillation interpretations of the event excess require additional sterile neutrino(s) with 
$\Delta m^2 \sim 1 $eV$^2/c^4$ (see \cite{oscill} for review). An alternative interpretation of the results of 
all three experiments is proposed in \cite{gninenko}. Below we briefly discuss
results obtained in that paper.

The main idea (proposed for the first time in \cite{gninenko2}) is that in the experiments mentioned above
signals from electrons and photons are indistinguishable. One could introduce heavy sterile
neutrino $\nu_h$ as a component of $\nu_\mu$ flavor eigenstate with a corresponding mixing
matrix element $U_{\mu h}$ which is produced in $\nu_\mu$ neutral current (NC) interactions and decays radiatively
into a photon and a light neutrino $\nu$. The decay channel $\nu_h\to\nu\gamma$ is dominant if there is a
large enough magnetic transition moment $\mu_{tr}$ (it requires substantial new physics
because in a minimally extended SM $\mu_{tr}$ is not large enough, see \cite{shrock1}).
In this case the event excess in LSND and MiniBooNE experiments comes from photons and not from electrons. 
In KARMEN experiment, $\nu_h$'s with $m >$ 40 MeV/c$^2$ cannot be produced within the detector
because of a kinematic threshold effect.

Sterile neutrino $\nu_h$ could be either of a Dirac or Majorana type. In the latter case a photon angular
distribution in  $\nu_h$ rest frame is isotropic while for the Dirac case there is an anisotropy depending
 on $\nu_h$ mass: $\myfrac{dN}{dcos\theta^\star} \sim (1 +  \myfrac{m_\mu^2 - m_h^2}{m_\mu^2 + m_h^2} cos\theta^\star)$.

The combined analysis of LSND, KARMEN and MiniBooNE data results in the following properties of $\nu_h$
(regardless of the neutrino type): 
\begin{itemize}
\item 40MeV/c$^2 \lesssim m_h \lesssim$ 80MeV/c$^2$;
\item $10^{-11}s \lesssim \tau_h \lesssim 10^{-9}$s;
\item $10^{-3} \lesssim |U_{\mu h}|^2 \lesssim 10^{-2}$.
\end{itemize}

It was mentioned that $\nu_h$ could be considered as a component of $\nu_\mu$.
 This leads to a very important consequence that $\nu_h$ is also
produced in charged current (CC) interactions and 
can be effectively searched for in pion and kaon decays (this idea was proposed in \cite{shrock2}).

The simplest way to do it is to study two-body decays $\pi\to\mu\nu$ 
and $K\to\mu\nu$  and look for a peak in the muon energy distribution 
($ E_\mu = (M^2 + m_\mu^2 - m_h^2)/2M$)
below the main one from
$\pi\to\mu\nu_\mu$ ($\pi_{\mu2}$) and $K\to\mu\nu_\mu$ ($K_{\mu2}$).
These decays allow to search for $\nu_h$ with masses up to $\sim$300 MeV/c$^2$.

Experimental limits from $\pi_{\mu2}$ decay \cite{pimu2_hnu} were obtained for 5MeV/c$^2 \le m_h \le$ 30MeV/c$^2$: $|U_{\mu h}|^2 < 10^{-5} - 10^{-3}$.
Best limits for kaon decays come from KEK experiment \cite{kmu2_hnu}: $|U_{\mu h}|^2 < 10^{-4}$ for 70MeV/c$^2 \le m_h \le$ 300MeV/c$^2$.
$\pi_{\mu2}$ decay is not sensitive to large $m_h$ masses, while
$K_{\mu2}$ decay is not sensitive to low $m_h$ masses because of resolution effects and strong background from 
$K\to\mu\nu_\mu\gamma (K_{\mu2\gamma}$) decay. Thus, a region 30MeV/c$^2 < m_h <$ 70MeV/c$^2$ is not constrained at all.
One should notice that limits above were obtained for relatively long-lived neutrinos
flying away from a detector (photon veto was applied in both cases).

Another possibility to search for heavy neutrino in kaon decays 
(which we are going to use)
is to measure $K\to\mu\nu_h (\nu_h \to \nu \gamma)$
decay chain. In this case the background from $K_{\mu2}$ is small and one can search for $\nu_h$ in a low mass region
(the background from $K_{\mu2\gamma}$ is also small, see Section \ref{signal_extraction}).
One should stress here that only the case of radiatively decayed neutrinos is considered.

The main purpose of this Letter is to search for heavy neutrino in $K^-\to\mu^-\nu_h (\nu_h \to \nu \gamma)$
with the properties described in \cite{gninenko} and in the following parameter range: 
30MeV/c$^2 \le m_h \le$ 80MeV/c$^2$, $10^{-11}s \le \tau_h \le 10^{-9}$s.

\section{ISTRA+ setup}

\subsection{Experimental setup}
\begin{center}
\begin{figure}[h]
\includegraphics[scale=.55 , angle=90]{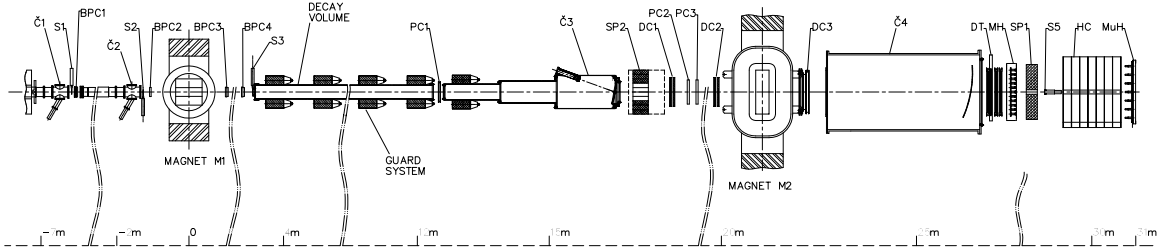}
\caption{ Elevation view of the ISTRA+ detector.}\label{detector}
\end{figure}
\end{center}
The experiment was performed at the IHEP 70 GeV proton synchrotron U-70.
The experimental setup ISTRA+ (fig.~\ref{detector}) was described in details
in \cite{ISTRA}. 
 The setup was located in the  negative unseparated secondary beam. 
The beam momentum in the measurements was $\sim 26$ GeV/c with 
$\Delta p/p \sim 1.5 \% $. The fraction of $K^{-}$ in the beam was $\sim 3 \%$.
The beam intensity was $\sim 3 \cdot 10^{6}$ per 1.9 s U-70 spill.
The track of a beam particle deflected by the magnet $M_{1}$ was 
measured by $BPC_{1}- BPC_{4}$ (1mm step multiwire chambers), the kaon identification was done by 
$\check{C_{0}} - \check{C_{2}}$ threshold Cherenkov counters. 
A 9 meter long vacuum
decay volume was surrounded by the Guard System ($GS$) -- 8 lead glass rings $LG_{1} - LG_{8}$ used to
veto low energy photons. $SP_{2}$ was a lead glass calorimeter to detect/veto 
large angle photons.
Tracks of decay products deflected in $M_2$ with 1Tm field integral
were measured by $PC_{1} - PC_{3}$ (2mm step proportional chambers);
$DC_{1} - DC_{3}$ (1cm cell drift chambers) and finally by 2cm diameter
drift  tubes    $DT_{1} - DT_{4}$. 
Wide aperture threshold Cherenkov counters $\check{C_{3}}$, 
$\check{C_{4}}$ were filled  with He and
were not used in the measurements. Nevertheless $\check{C_{3}}$ was used as an extension of the decay volume.
$SP_{1}$ ($ECAL$) was a 576-cell lead glass calorimeter,
followed by $HC$ ($HCAL$) -- a scintillator-iron sampling hadron calorimeter. $HC$ was subdivided
into 7 longitudinal sections 7$\times$7 cells each. $MH$ was a 
11$\times$11 cell scintillating hodoscope used to  improve the time resolution of the 
tracking system, $MuH$  was a  7$\times$7 cell muon hodoscope. 

The trigger was provided by $S_{1} - S_{3}$, $S_5$ scintillation counters, 
$\check{C_{0}} - \check{C_{2}}$ Cherenkov counters,
the analog sum of amplitudes from the last dinodes of the $SP_1$ :
 $T_{0}=S_{1} \cdot S_{2} \cdot S_{3} \cdot 
 \check{C_{0}} \cdot \bar{\check{C_{1}}} \cdot 
 \bar{\check{C_{2}}} \cdot 
 \bar{S_{5}} \cdot \Sigma(SP_{1})$,
here  $S_5$ was a counter downstream  the setup at the beam focus;
$\Sigma(SP_{1})$ -- a requirement for the analog sum of $ECAL$ amplitudes 
to be above $\sim$3 GeV. The last requirement 
served to suppress the $K_{\mu2}$ decay.
About $10\%$ events were recorded with a different trigger:
 $T_{1}=S_{1} \cdot S_{2} \cdot S_{3} \cdot 
 \check{C_{0}} \cdot \bar{\check{C_{1}}} \cdot 
 \bar{\check{C_{2}}} \cdot 
 \bar{S_{5}}$.
This prescaled trigger allowed to calculate the trigger efficiency as a function of the energy released in $ECAL$.

\subsection{Data and MC samples}

We use high-statistics data collected in Winter 2001 run. About 332M events were stored on tapes.
This statistics was complemented by ~200M MC events generated with Geant3 \cite{geant3}. The MC generation includes a realistic description
of all ISTRA+ detectors.

For the signal simulation kaons are forced to decay into a muon and $\nu_h$.
A new particle $\nu_h$ is introduced to Geant3:
\begin{itemize}
\item $\nu_h$ mass  is 30, 40, 50, 60, 70 and 80 MeV/c$^2$;
\item $\nu_h$ lifetime is $10^{-9}$, $10^{-10}$ and $10^{-11}$s; 
\item $\nu_h$ decays into a photon and a massless neutrino;
\item photon angular distribution in $\nu_h$ rest frame is isotropic (Majorana case).
An anisotropy for the Dirac type of $\nu_h$ is obtained by weighting MC events.
\end{itemize}
Each sample of the signal MC (for particular values of $m_h$ and $\tau_h$) contains 1M event.

\section{Event reconstruction}

\subsection{Photon momentum reconstruction}

Event reconstruction for $K^-\to\mu^-\nu_h (\nu_h \to \nu \gamma)$ is nearly the same as for $K^-\to\mu^-\nu_\mu \gamma$.
The only difference is that for $K^-\to\mu^-\nu_\mu \gamma$ it is possible to reconstruct the gamma momentum in the laboratory frame $\vec p^{~lab}_\gamma$
using the decay vertex
and a shower centre while for $K^-\to\mu^-\nu_h (\nu_h \to \nu \gamma)$ the photon is emitted from a secondary vertex which is unknown.
Nevertheless one can reconstruct $\vec p^{~lab}_\gamma$ using a primary vertex.
 This leads to an additional photon energy smearing in the kaon rest frame.

\begin{figure}[!h]
\centering
\includegraphics[width=15cm , angle=0]{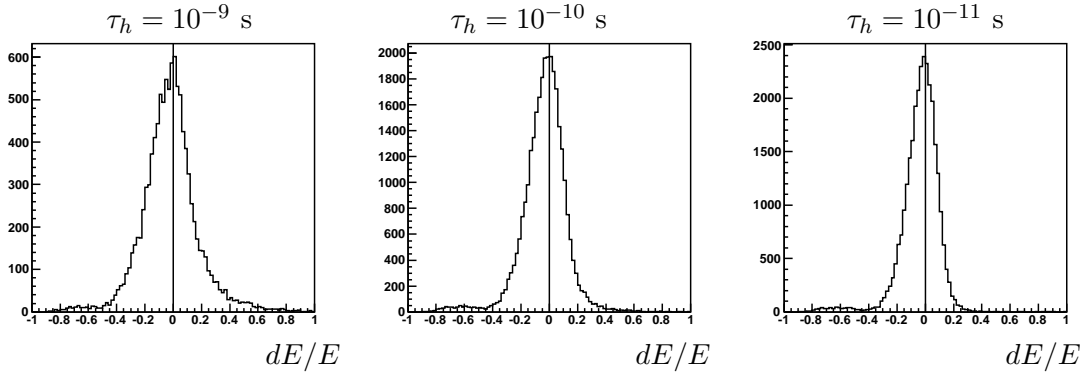}
\caption{Photon energy smearing in the kaon rest frame: $dE/E = (E_{measured} - E_{true})/E_{true}$ (signal MC, $m_h$=60MeV/c$^2$).}\label{de}
\begin{picture}(1,1)
\put(33,43){\small $dE/E$}
\put(-107,43){\small $dE/E$}
\put(173,43){\small $dE/E$}
\put(-167,170){\small $\tau_h=10^{-9}$ s}
\put(-37,170){\small $\tau_h=10^{-10}$ s}
\put(113,170){\small $\tau_h=10^{-11}$ s}
\end{picture}
\end{figure}

 In Fig.~\ref{de} the normalized difference between the measured and 
true photon energy in the kaon rest frame is shown.
For $\tau_h = 10^{-11} $s this difference is dominated by resolution effects while for large $\tau_h$ one can see an additional smearing.
Typical values of the photon transverse momentum with respect to $\nu_h$ momentum are small, that is why the smearing is not crucial.

\subsection{Signal signatures for $K\to\mu\nu_h(\nu_h\to\nu\gamma)$}
\label{sec_4}

The simplest way to observe heavy neutrino is to look for a peak in $E_\mu$ -- muon energy in the kaon rest frame
(see Section \ref{sec_one}).

Another signature comes from a photon angular distribution in the kaon rest frame. For this distribution
using Lorentz boost transformation one can obtain the following formula:
\begin{center}
$\myfrac{dN}{dcos\theta} = \myfrac{dN}{dcos\theta^\star} \cdot \myfrac{1}{\gamma^2(\beta cos\theta-1)^2}$ 
\end{center}
where $\theta$ is the angle between $\vec{p_h}$ and $\vec{p}_\gamma$ in the kaon rest frame,  
$\theta^\star$ is the angle between $\vec{p}_\gamma^{~\star}$ in $\nu_h$ rest frame and the boost axis (along $\vec{p_h}$),
$\gamma = \myfrac{E_h}{m_h}$. 
The term $\myfrac{dN}{dcos\theta^\star}$ is constant for the Majorana type of $\nu_h$. In the Dirac case 
$\myfrac{dN}{dcos\theta^\star} \sim (1 +  \myfrac{m_\mu^2 - m_h^2}{m_\mu^2 + m_h^2} cos\theta^\star)$.

In all cases $\gamma > 3$, therefore $\beta \sim 1$ and the angular distribution has a peak at $cos\theta \sim 1$ and~
~hence $cos\theta_{\mu\gamma} \sim -1$ in the kaon rest frame. This peak 
is a very good signature for the signal.

\begin{figure}[!h]
\centering
\includegraphics[width=10cm , angle=0]{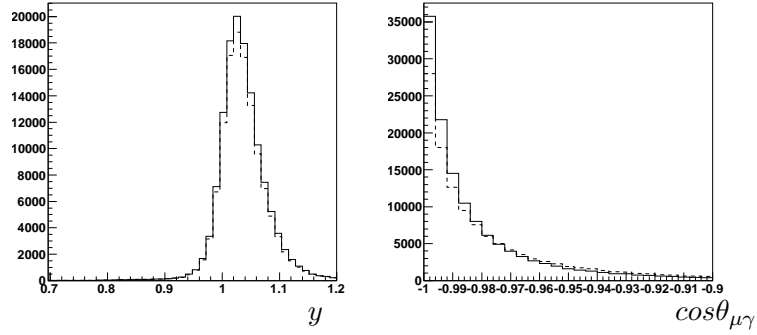}
\caption{Dirac (solid) and Majorana (dashed) type of $\nu_h$ ($m_h$=60MeV/c$^2$, $\tau_h=10^{-10}s)$.
Left: distribution over normalized muon energy
$y = 2E_\mu/m_K$
(lines - selected region, see Section \ref{sel_region}). Right: $cos\theta_{\mu\gamma}$ in the kaon rest frame.
}\label{w_dirac}
\begin{picture}(1,1)
\put(-27,60){\small $y$}
\put(112,60){\small $cos\theta_{\mu\gamma}$}
\end{picture}
\end{figure}

 Distributions over $cos\theta_{\mu\gamma}$ and $y$ for Dirac and Majorana cases are shown in Fig. \ref{w_dirac}.
It can be seen that the difference in $y$ and $cos\theta_{\mu\gamma}$ shapes is negligible for two neutrino types.

\subsection{Primary and secondary decay vertex}

\begin{figure}[!h]
\centering
\includegraphics[width=15cm , angle=0]{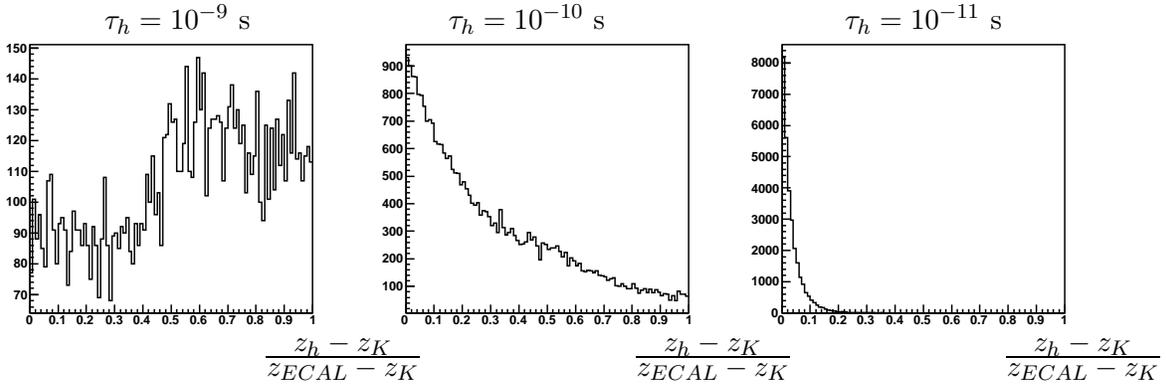}
\caption{The difference between coordinates of the secondary ($z_h$) and primary ($z_K$) vertices
 divided by the distance from $z_K$ to $ECAL$ (MC, $m_h$=60MeV/c$^2$).}\label{dz}
\begin{picture}(1,1)
\put(33,43){\small $\myfrac{z_h - z_K}{z_{ECAL} - z_K}$}
\put(-107,43){\small $\myfrac{z_h - z_K}{z_{ECAL} - z_K}$}
\put(173,43){\small $\myfrac{z_h - z_K}{z_{ECAL} - z_K}$}
\put(-167,170){\small $\tau_h=10^{-9}$ s}
\put(-37,170){\small $\tau_h=10^{-10}$ s}
\put(113,170){\small $\tau_h=10^{-11}$ s}
\end{picture}
\end{figure}

The difference between $z$-coordinates ($z$-axis is collinear with the beam direction) 
of the secondary and primary vertices divided 
by the distance between the primary vertex and the electromagnetic calorimeter
is shown in Fig.~\ref{dz}. A distribution over this ratio shows a fraction of $\nu_h$ decays within the decay volume.
For $\tau_h = 10^{-11}$s and $\tau_h = 10^{-10}$s almost all neutrinos decay within an experimental setup, while
 for $\tau_h = 10^{-9}$s a geometrical inefficiency becomes substantial.

\section{Event selection}
\label{sec_6}

The event selection for $K^-\to\mu^-\nu_h (\nu_h \to \nu \gamma)$ is very similar to that of $K^-\to\mu^-\nu_\mu \gamma$. 
Standard kinematic variables are used for the further analysis: $x=2E_\gamma/m_K$ and $y=2E_\mu/m_K$, 
$E_\gamma$ and $E_\mu$ being photon and muon energies in the kaon rest frame. As in \cite{istra_mng}, Dalitz-plot will be used
for studying signal and background kinematic regions.

The decay signature is defined as follows: one primary track (kaon);
 one negatively charged secondary track identified as muon;
 one shower in $ECAL$ not associated with the charged track. Muon identification 
using $ECAL$ and $HCAL$ is described in our previous papers (\cite{kmu3-1,kmu3-2}).

Several cuts are applied to clean the data:
\begin{itemize}
\item number of beam and decay track projections in $XZ$ and $YZ$ planes is equal to 1;
\item CL (confidence level of a track fit) for the beam track projections in both planes must be greater 
 than 10$^{-2}$;
\item CL for the decay track projections is greater than 0.1 ($XZ$) and 0.15 ($YZ$);
\item the angle between a primary (kaon) and secondary (muon) track is greater than 2~mrad. 
\end{itemize}
The last cut eliminates most of undecayed beam particles.
The quality of the decay track (described quantitatively by CL) is worse than that of the beam track because of multiple scattering and detector resolution.

Cuts containing a photon energy include:
\begin{itemize}
\item gamma energy in the kaon rest frame is greater than 10 MeV;
\item no photons in $SP_2$ calorimeter (the energy threshold is 0.5 GeV for the total energy release);
\item no photons in $GS$.
\end{itemize}

For vertex characteristics we have the following requirements:
\begin{itemize}
\item $z$-coordinate must be within the interval 400 $<$ z$_{vtx}$ $<$ 1600cm;
\item  (-3) $< x_{vtx} <$ 3cm;
\item  (-2) $< y_{vtx} <$ 6cm;
\item CL of general vertex fit is greater than 10$^{-2}$.
\end{itemize}

Additional cuts are applied to suppress backgrounds:
\begin{itemize}
\item number of hits in the matrix hodoscope ($MH$) is less than 3;
\item missing momentum $\overrightarrow p^{lab}_{miss}=\overrightarrow p^{lab}_K-\overrightarrow p^{lab}_\mu-\overrightarrow p^{lab}_\gamma$ 
does not point to the $ECAL$ central hole (this cut effectively rejects the background from $K^-\to\pi^-\pi^0$ decay when
 the lost photon from $\pi^0\to\gamma\gamma$ goes into the hole).
\end{itemize}

\subsection{Trigger efficiency}
\begin{figure}[h]
\centering
\includegraphics[width=6cm , angle=0]{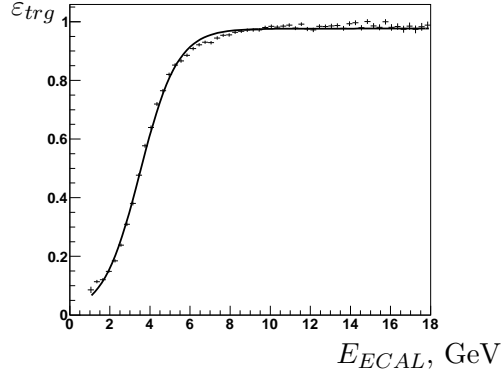}
\caption{$T_0$ trigger efficiency. Points -- data, curve -- fit by the Fermi function.}\label{trig}
\begin{picture}(1,1)
\put(33,33){\small $E_{ECAL}$, GeV}
\put(-90,165){\small $\varepsilon_{trg}$}
\end{picture}
\end{figure}

As $T_0$ trigger described in Section 2 contains the energy threshold in $SP_1$, the trigger efficiency 
as a function of energy released in $ECAL$ should be known. It could be found using events with $T_1$ trigger:
$\varepsilon_{trg}=(T_1\bigcap T_0)~/~ T_1$. The trigger curve is shown in Fig.~\ref{trig}. The fit is done using a Fermi function.
For the further analysis only events with $T_0$ are kept and these events are weighted by the factor of $1/\varepsilon_{trg}$.

\section{Signal extraction}
\label{signal_extraction}

\begin{figure}[h]
\begin{minipage}[t]{0.45\textwidth}
\centering
\includegraphics[width=5.5cm , angle=0]{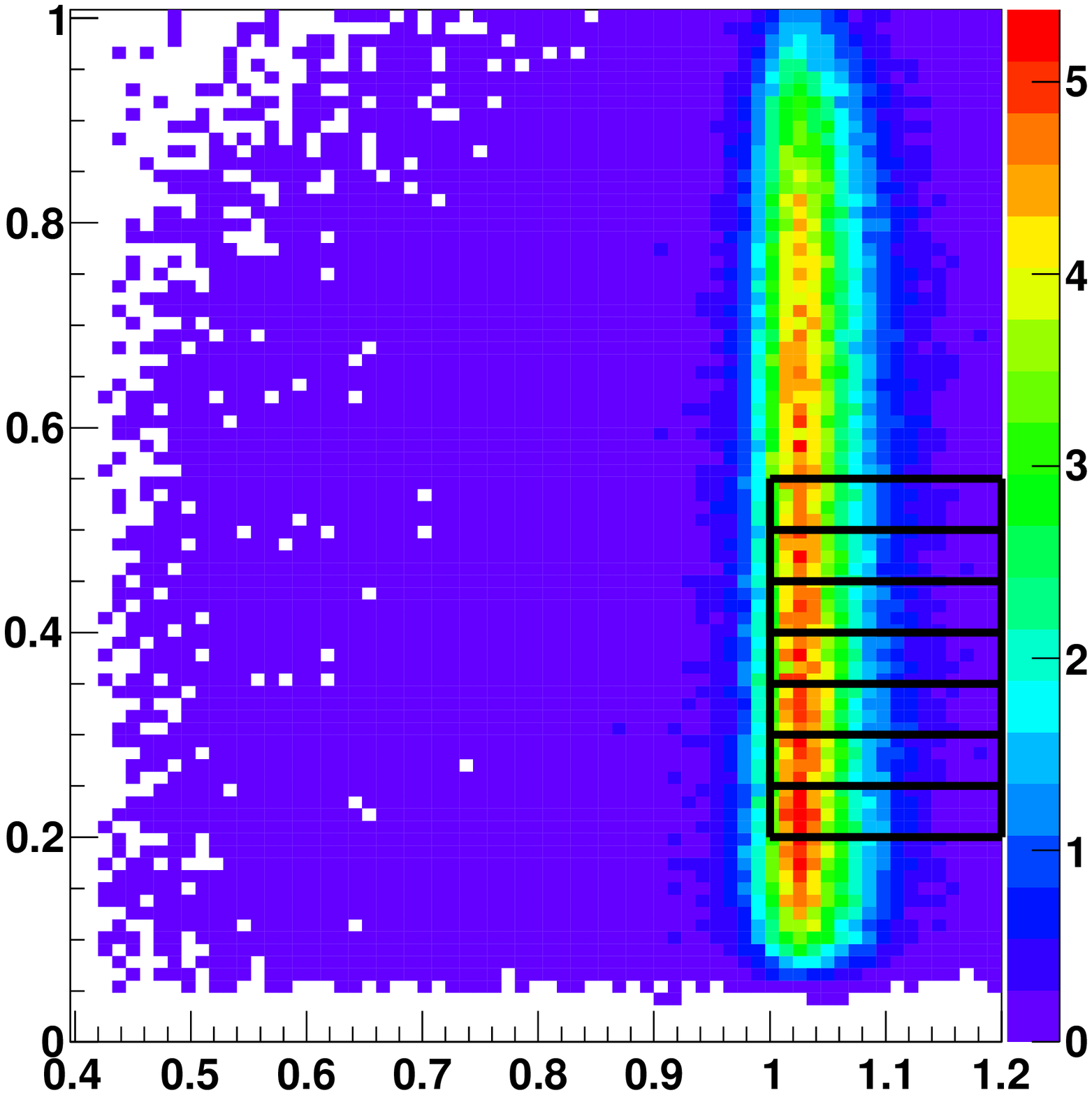}
\caption{Dalitz-plot density for the signal ~~~~~~~~~~~~~~~~~~~~~~~~~($m_h$=60MeV/c$^2$, $\tau_h=10^{-10}s)$.}\label{hnu_dalitz}
\end{minipage}  
\hspace{1cm}
\begin{minipage}[t]{0.45\textwidth}
\centering
\includegraphics[width=5.5cm , angle=0]{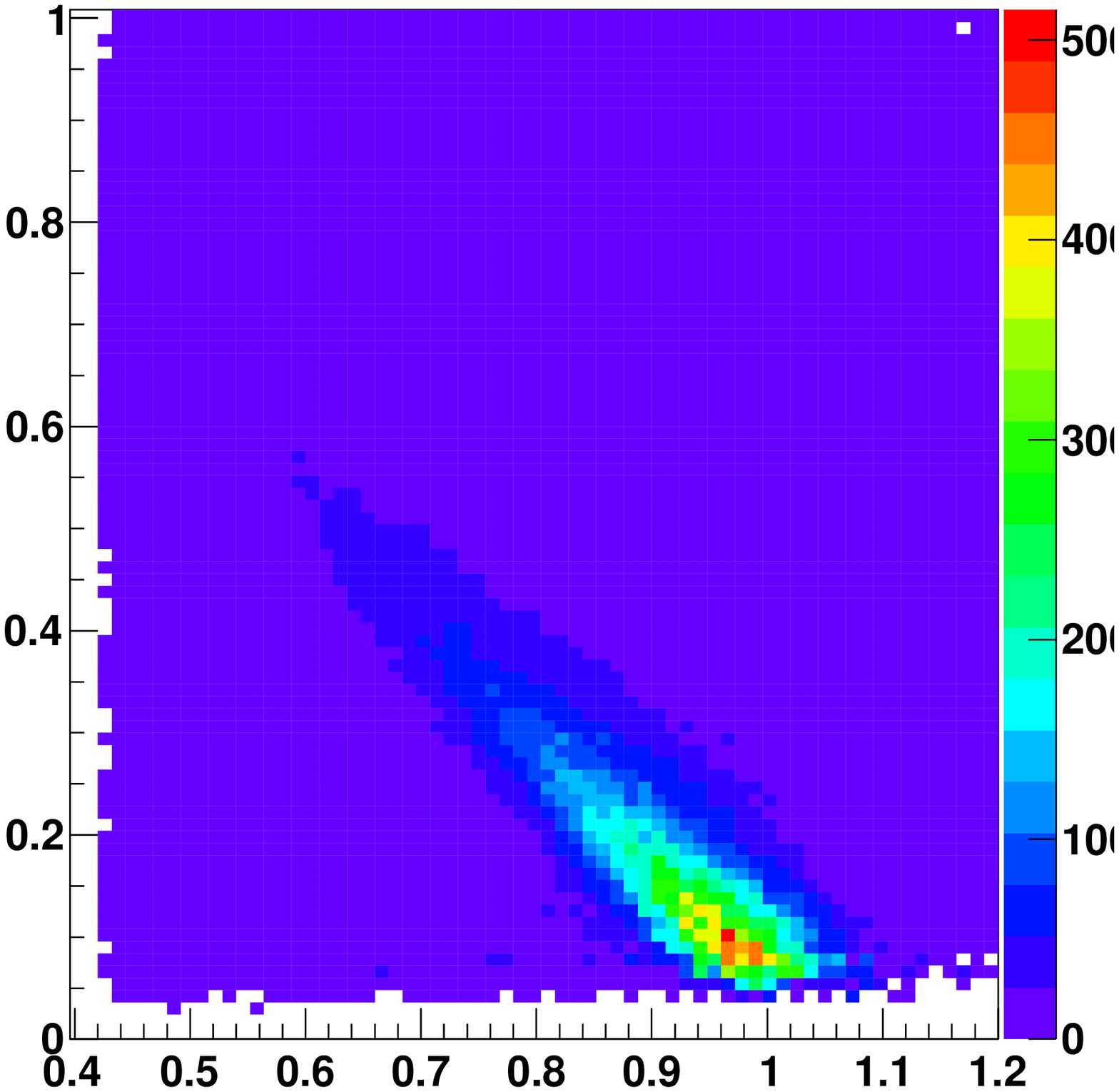}
\caption{Dalitz-plot density for the $K_{\mu2\gamma} (IB) $ background.}\label{kmu2g_dalitz}
\end{minipage}
\begin{picture}(1,1)
\put(-50,0){$y$}
\put(-290,0){$y$}
\put(-180,140){$x$}
\put(-420,140){$x$}
\end{picture}
\end{figure}

\begin{figure}[h]
\begin{minipage}[t]{0.45\textwidth}
\centering
\includegraphics[width=5.5cm , angle=0]{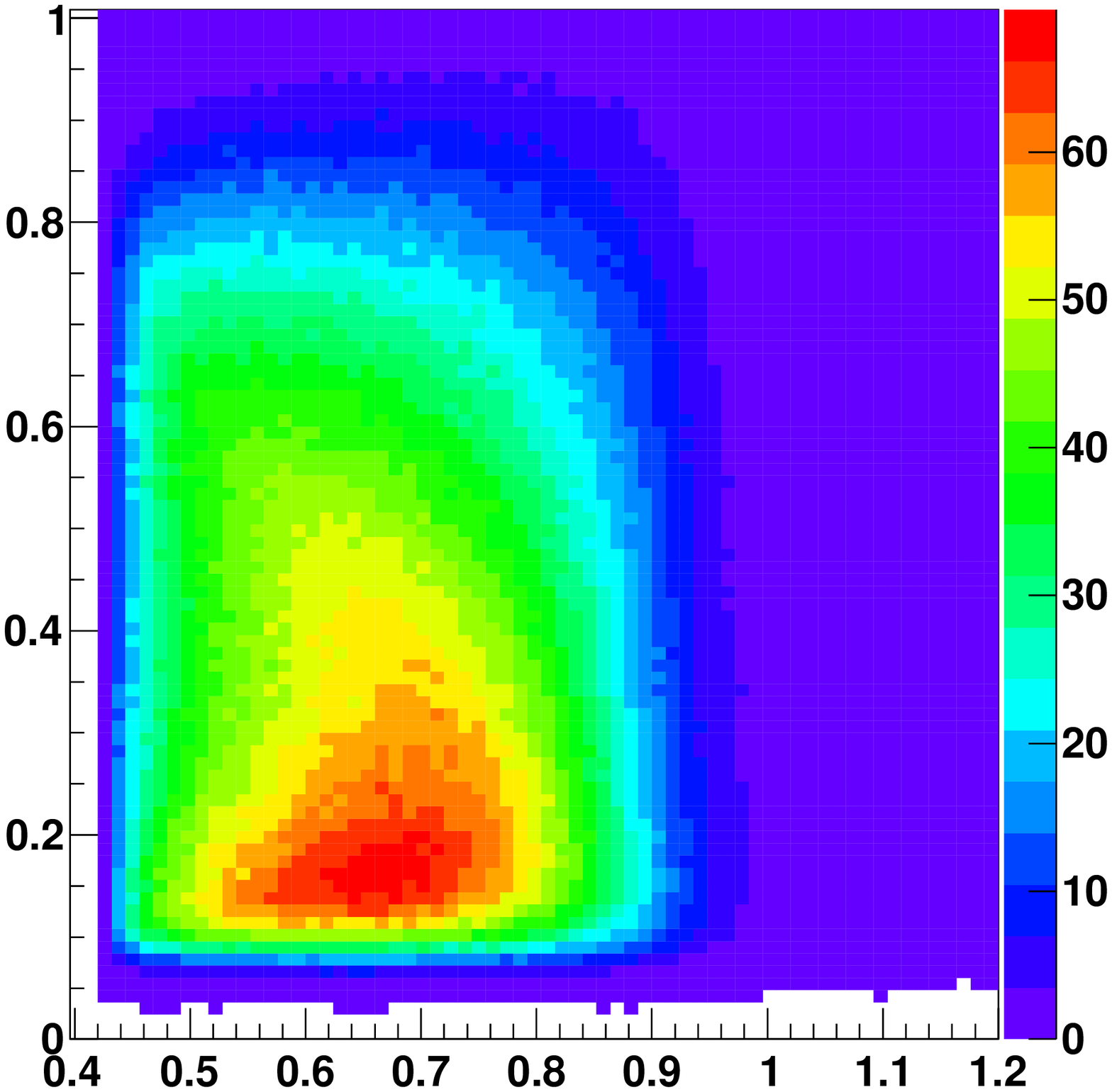}
\caption{Dalitz-plot density for the $K_{\mu3}$ background.}\label{bkgr1}
\end{minipage} 
 \hspace{1cm}
\begin{minipage}[t]{0.45\textwidth}
\centering
\includegraphics[width=5.5cm , angle=0]{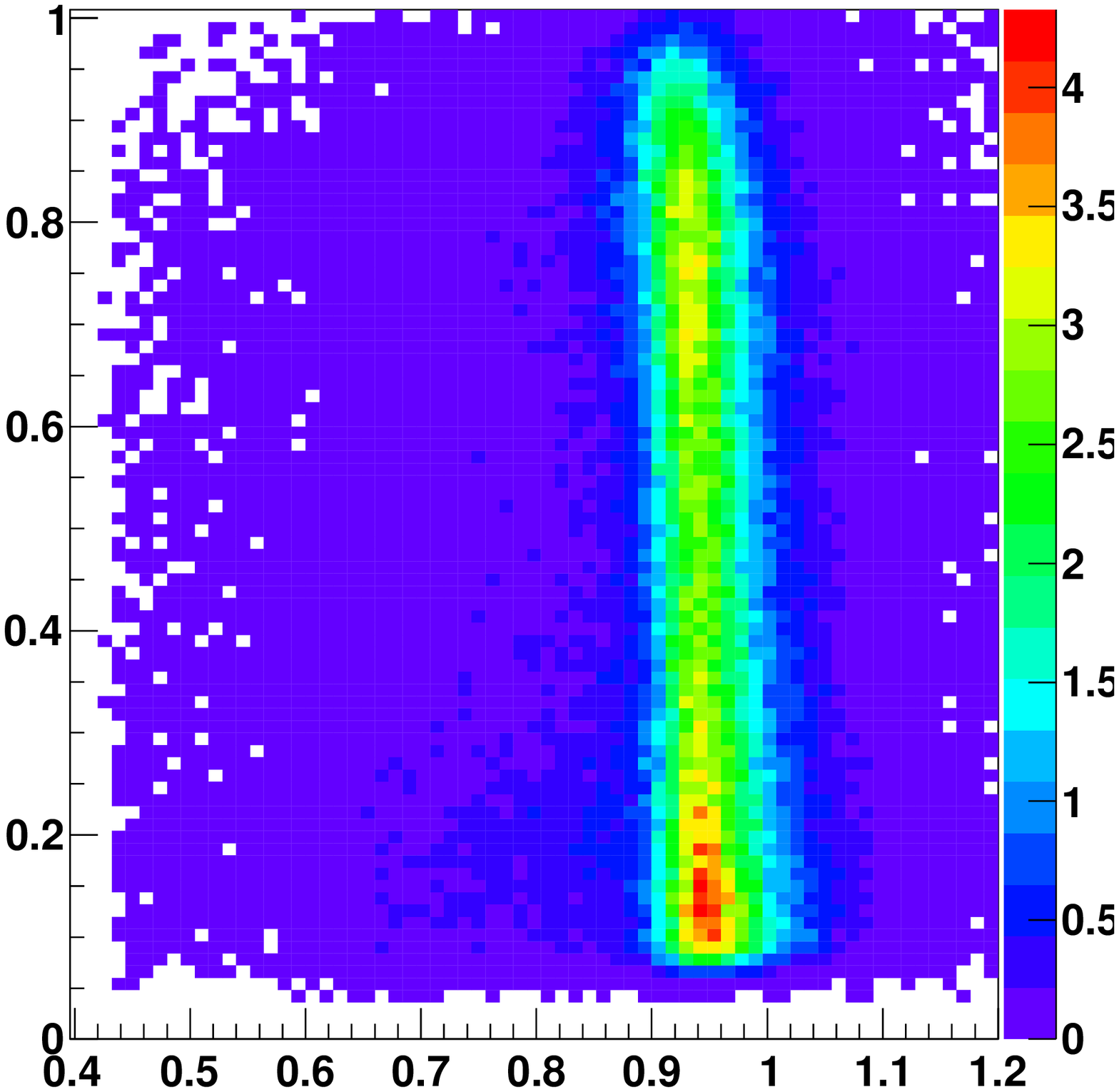}
\caption{Dalitz-plot density for the $K_{\pi2}$ background.}\label{bkgr2}
\end{minipage}
\begin{picture}(1,1)
\put(-50,0){$y$}
\put(-290,0){$y$}
\put(-180,140){$x$}
\put(-420,140){$x$}
\end{picture}
\end{figure}

As it was mentioned in Section~\ref{sec_6}, Dalitz-plot is used for signal and backgrounds studies.
The main background comes from 3 decay modes: $K^-\to\mu^-\bar\nu_\mu\gamma (K_{\mu2\gamma})$, $K^-\to\mu^-\bar\nu_\mu\pi^0 (K_{\mu3})$
with one gamma lost from $\pi^0\to\gamma\gamma$
 and $K^-\to\pi^-\pi^0 (K_{\pi2})$ with one gamma lost 
 and $\pi$ misidentified as $\mu$. Dalitz-plot distributions for the signal, $K_{\mu2\gamma}$, $K_{\mu3}$ and $K_{\pi2}$ are 
shown in Figs.~\ref{hnu_dalitz} -- \ref{bkgr2}.

\subsection{Signal extraction procedure}

The procedure starts with dividing all the kinematic ($y,x$) region into stripes on $x$ ($x$-stripes). The $x$-stripe 
width is $\Delta$x=0.05 ($\Delta E_{\gamma}\sim 12$MeV).
In every $x$-stripe we put a cut on $y$: 
$1.0 < y < 1.2$ to suppress backgrounds. 

In Section \ref{sec_4} two signal signatures were described -- peaks in $y$ and $cos\theta_{\mu\gamma}$.
For each $x$-stripe we do a simultaneous fit of two histograms -- $cos\theta_{\mu\gamma}$ (with the cut on $y$ introduced above)
and $y$ (without this cut).

\subsection{Selected kinematic region}
\label{sel_region}

\begin{figure}[!h]
\centering
\includegraphics[width=5.cm , angle=0]{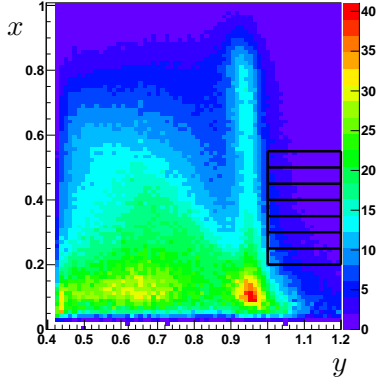}
\caption{Dalitz-plot density for the data.}\label{exp_dalitz}
\begin{picture}(1,1)
\put(53,38){\small $y$}
\put(-70,165){\small $x$}
\end{picture}
\end{figure}

For the further analysis we have selected seven $x$-stripes in the following region: 
$0.2 < x < 0.55$ (49MeV $< E_\gamma <$ 136MeV). Selected $x$-stripes are shown in Fig.~\ref{hnu_dalitz}.
Dalitz-plot for the data with selected $x$-stripes is shown in Fig.~\ref{exp_dalitz}.

\subsection{Possible signature for different $m_h$ and $\tau_h$}


\begin{figure}[!h]
\begin{minipage}[t]{0.3\textwidth}
\centering
\includegraphics[width=5cm , angle=0]{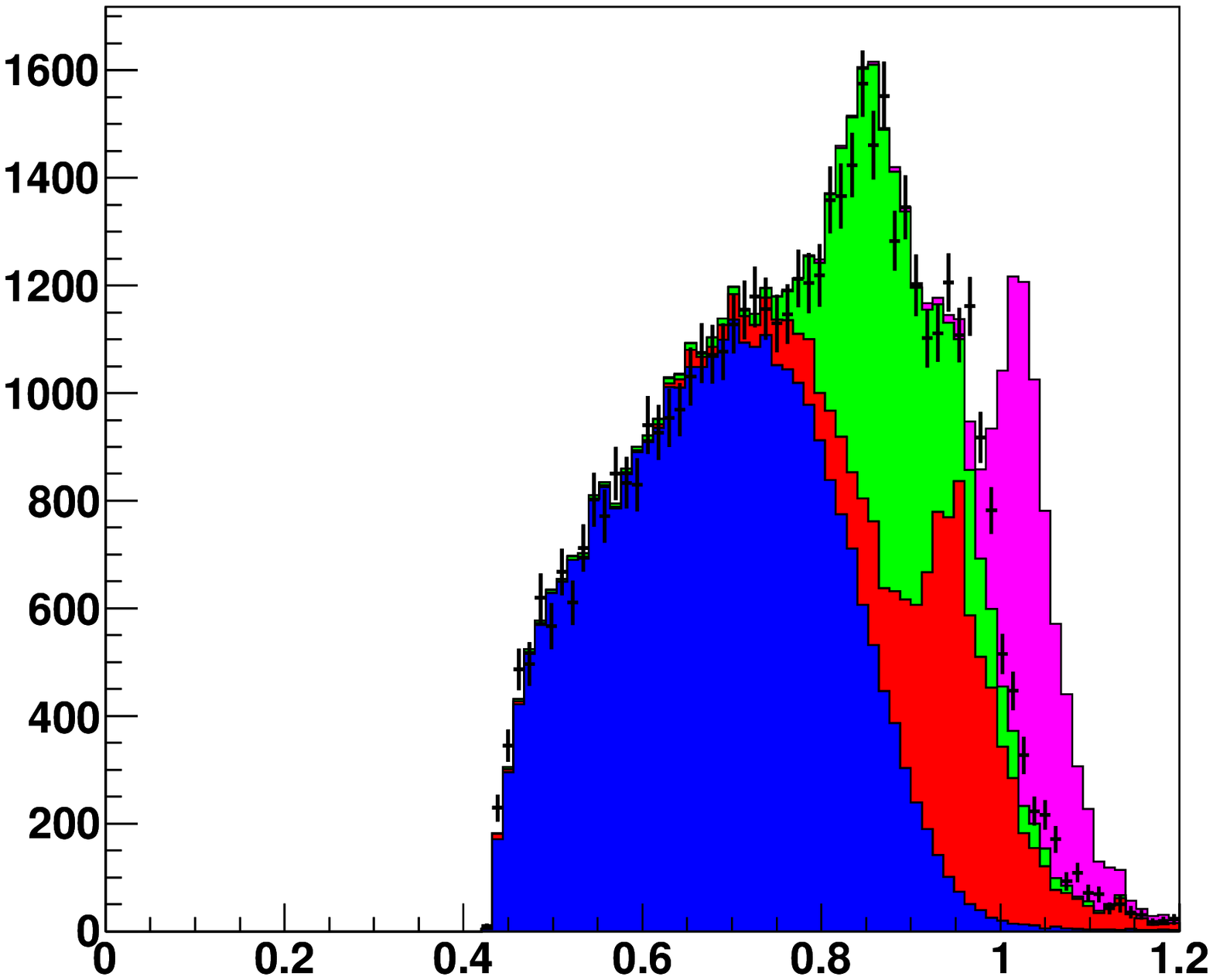}
\end{minipage}  
\hspace{0.2cm}
\begin{minipage}[t]{0.3\textwidth}
\centering
\includegraphics[width=5cm , angle=0]{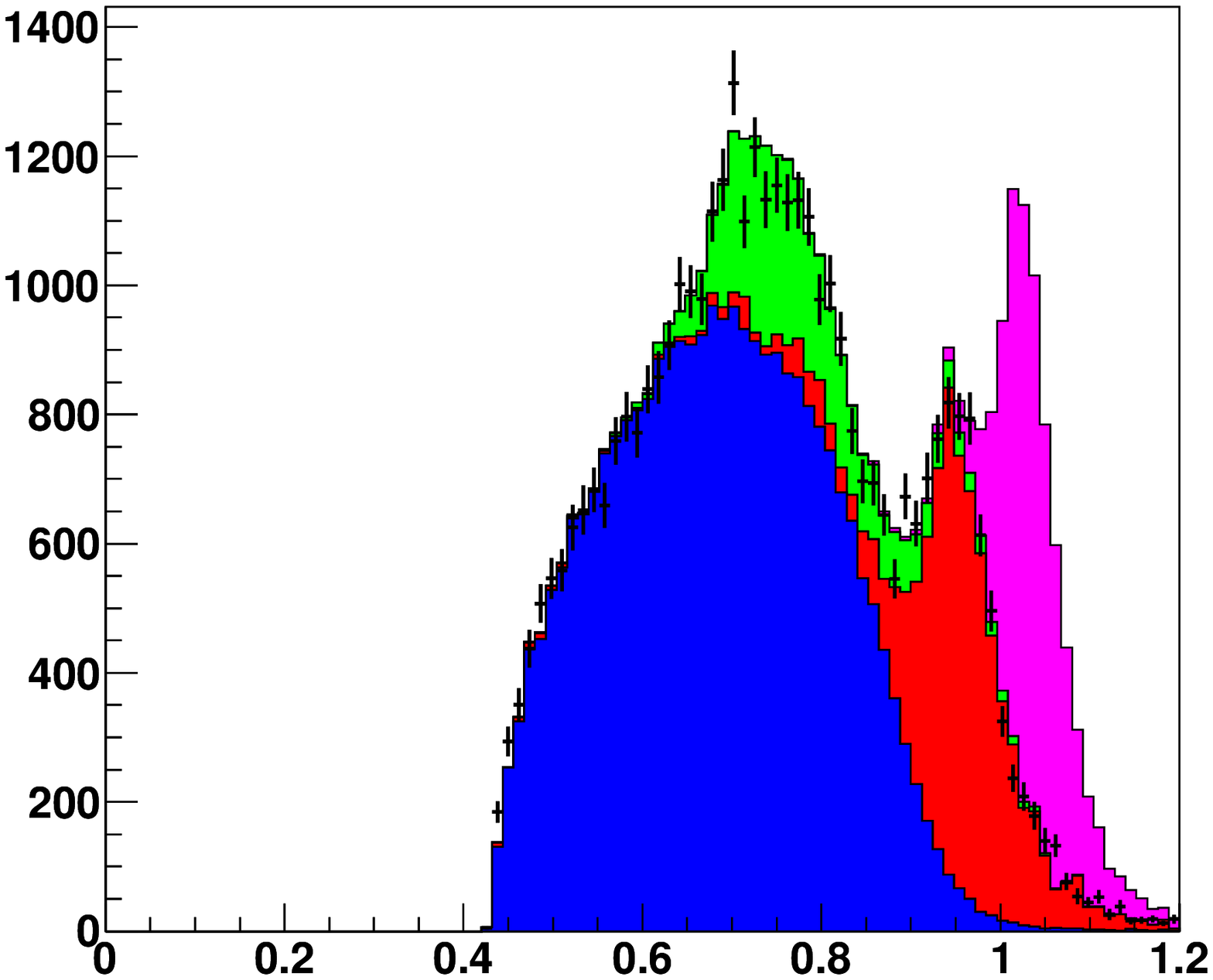}
\end{minipage}
\hspace{0.2cm}
\begin{minipage}[t]{0.3\textwidth}
\centering
\includegraphics[width=5cm , angle=0]{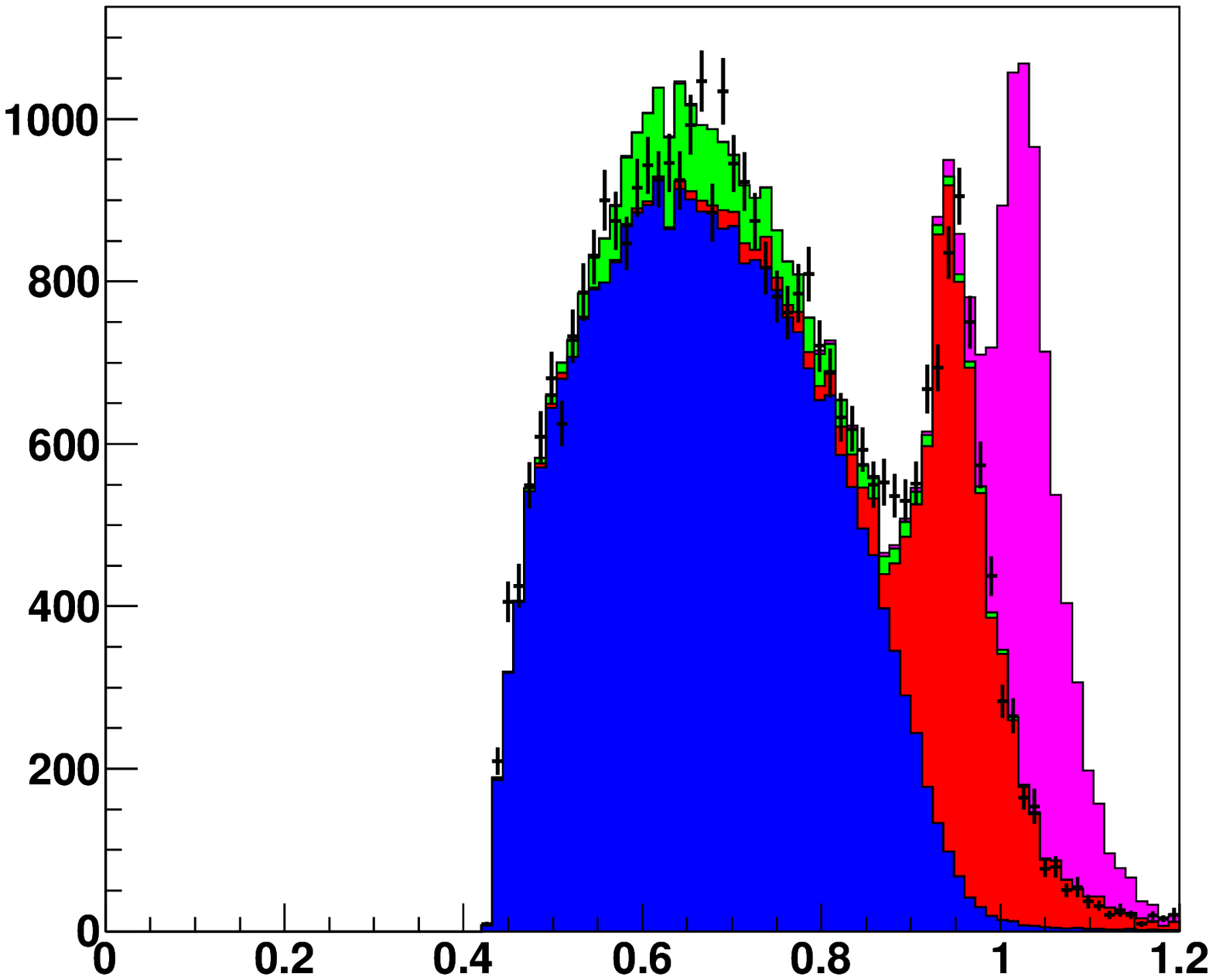}
\end{minipage} 
\caption{Distribution over $y$. Points with errors -- data, magenta histogram -- signal ($m_h$ = 60MeV/c$^2, \tau_h = 10^{-10}$s, $|U_{\mu h}|^2 = 10^{-2}$), red --  $K_{\pi2}$, blue -- $K_{\mu3}$ and green -- $K_{\mu2\gamma}$.}\label{y}
\begin{picture}(1,1)
\put(115,47){$y$}
\put(265,47){$y$}
\put(412,47){$y$}
\put(15,162){\footnotesize Stripe 1 ($0.2 < x < 0.25$)}
\put(165,162){\footnotesize Stripe 4 ($0.35 < x < 0.4$)}
\put(315,162){\footnotesize Stripe 7 ($0.5 < x < 0.55$)}
\end{picture}
\end{figure}

\begin{figure}[!h]
\begin{minipage}[t]{0.3\textwidth}
\centering
\includegraphics[width=5cm , angle=0]{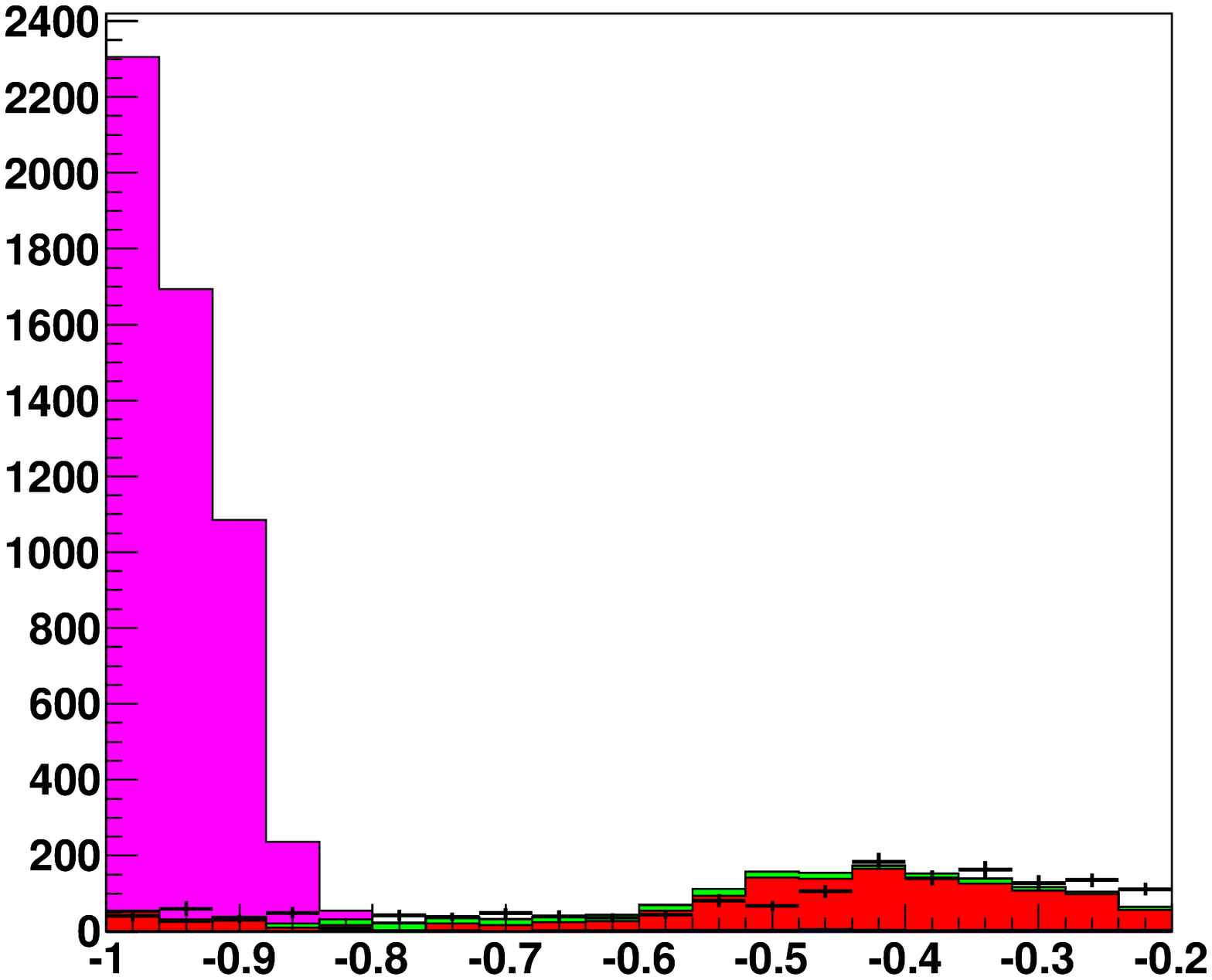}
\end{minipage}
\hspace{0.2cm}
\begin{minipage}[t]{0.3\textwidth}
\centering
\includegraphics[width=5cm , angle=0]{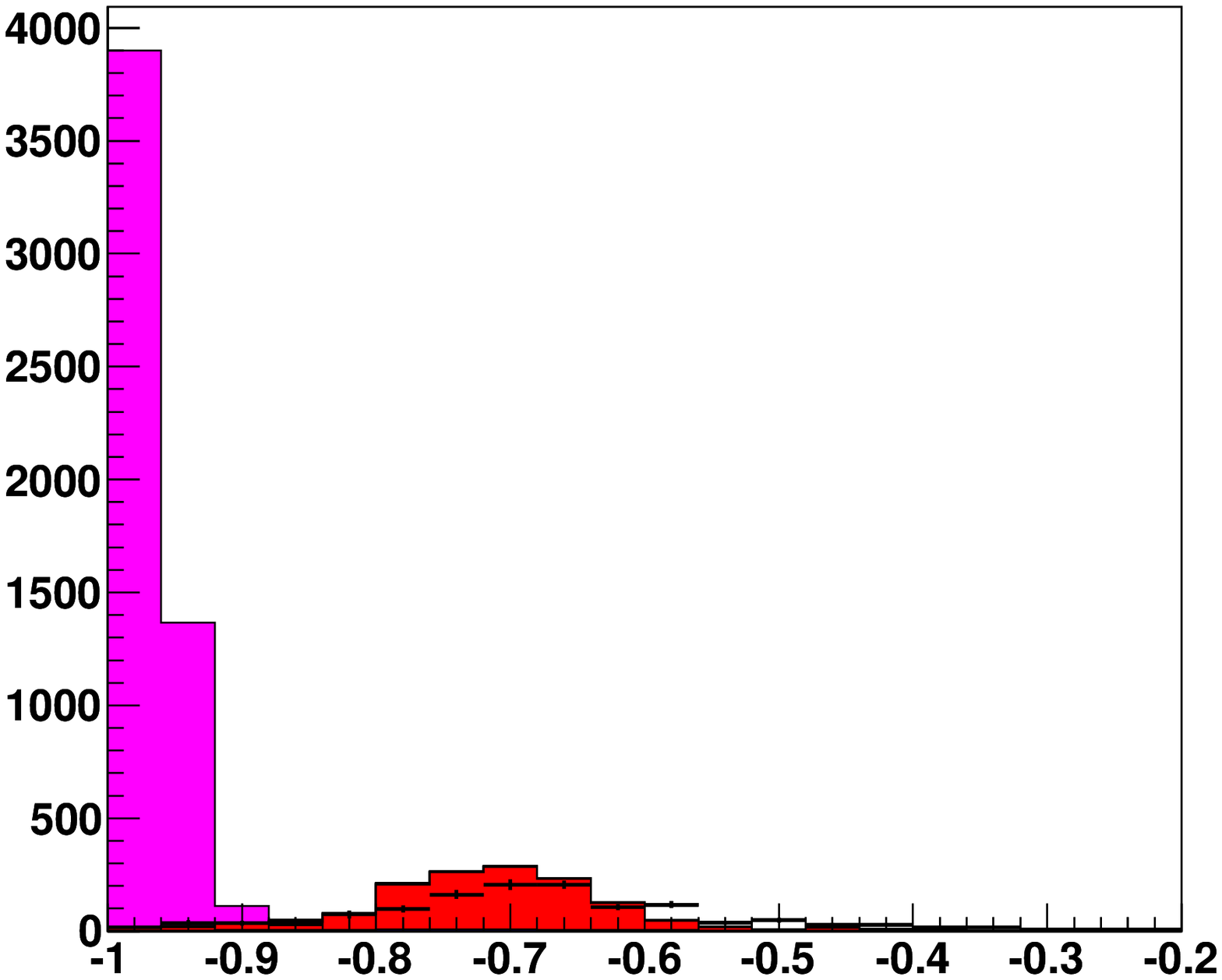}
\end{minipage}
\hspace{0.2cm}
\begin{minipage}[t]{0.3\textwidth}
\centering
\includegraphics[width=5cm , angle=0]{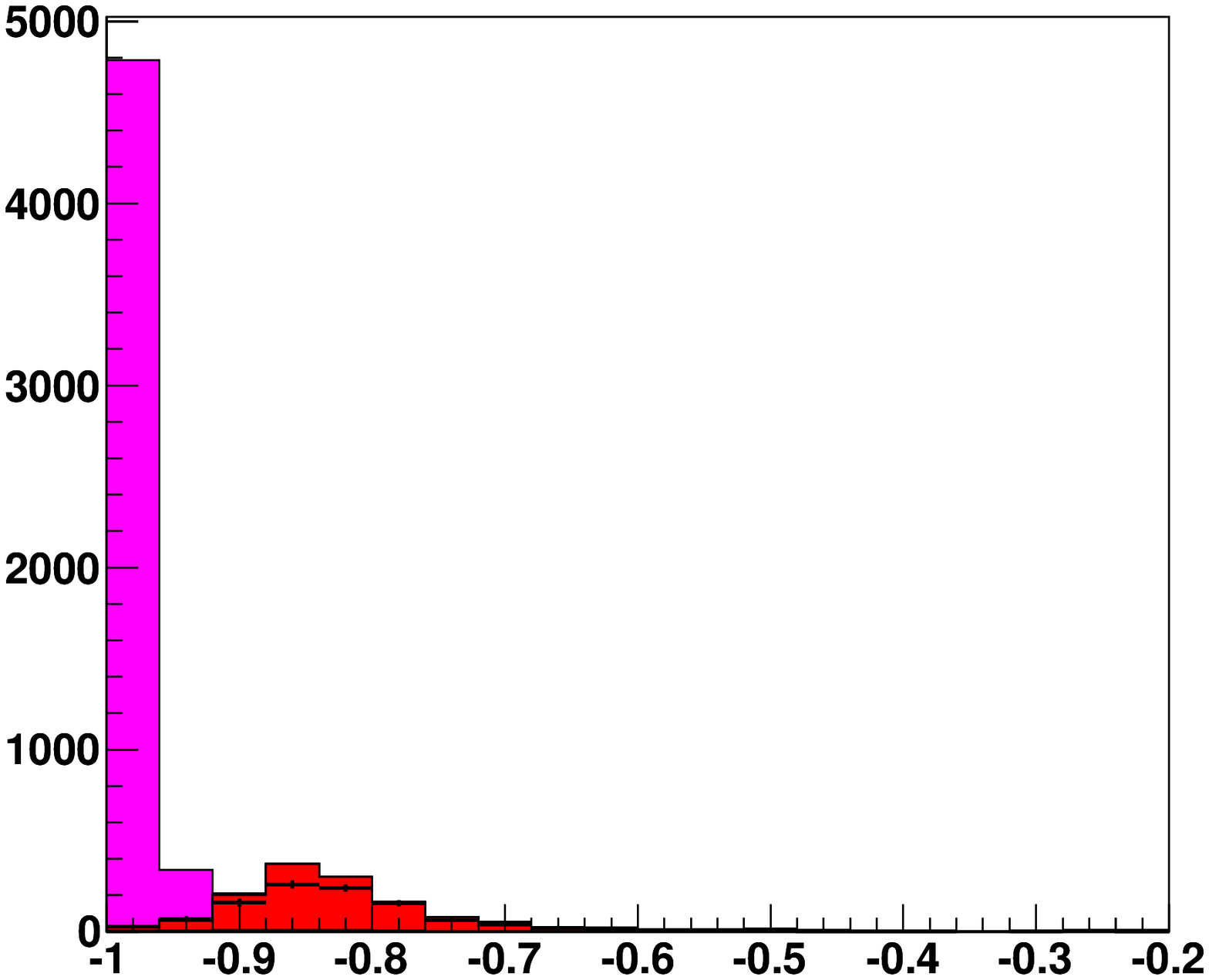}
\end{minipage} 
\caption{Distribution over $cos\theta_{\mu\gamma}$. Points with errors -- data, magenta histogram -- signal ($m_h$ = 60MeV/c$^2, \tau_h = 10^{-10}$s, $|U_{\mu h}|^2 = 10^{-2}$), red --  $K_{\pi2}$, blue -- $K_{\mu3}$ and green -- $K_{\mu2\gamma}$.}\label{cos}
\begin{picture}(1,1)
\put(115,47){$cos\theta_{\mu\gamma}$}
\put(265,47){$cos\theta_{\mu\gamma}$}
\put(412,47){$cos\theta_{\mu\gamma}$}
\put(15,162){\footnotesize Stripe 1 ($0.2 < x < 0.25$)}
\put(165,162){\footnotesize Stripe 4 ($0.35 < x < 0.4$)}
\put(315,162){\footnotesize Stripe 7 ($0.5 < x < 0.55$)}
\end{picture}
\end{figure}

To better understand how the signal looks like and how sensitive our data are to heavy neutrino we plot
$y$ and $cos\theta_{\mu\gamma}$ distributions for $m_h$ = 60MeV/c$^2$, $\tau_h = 10^{-10}$s and $|U_{\mu h}|^2 = 10^{-2}$
in three $x$-stripes (Figs.~\ref{y},~\ref{cos}). 
It can be seen that $cos\theta_{\mu\gamma}$ is much better for the signal observation
(or setting upper limits) while
$y$ is good for the reliable background normalization. 


\subsection{Simultaneous fit results}

\begin{center}
\begin{figure}[!h]
\centering
\includegraphics[scale=.5 , angle=0]{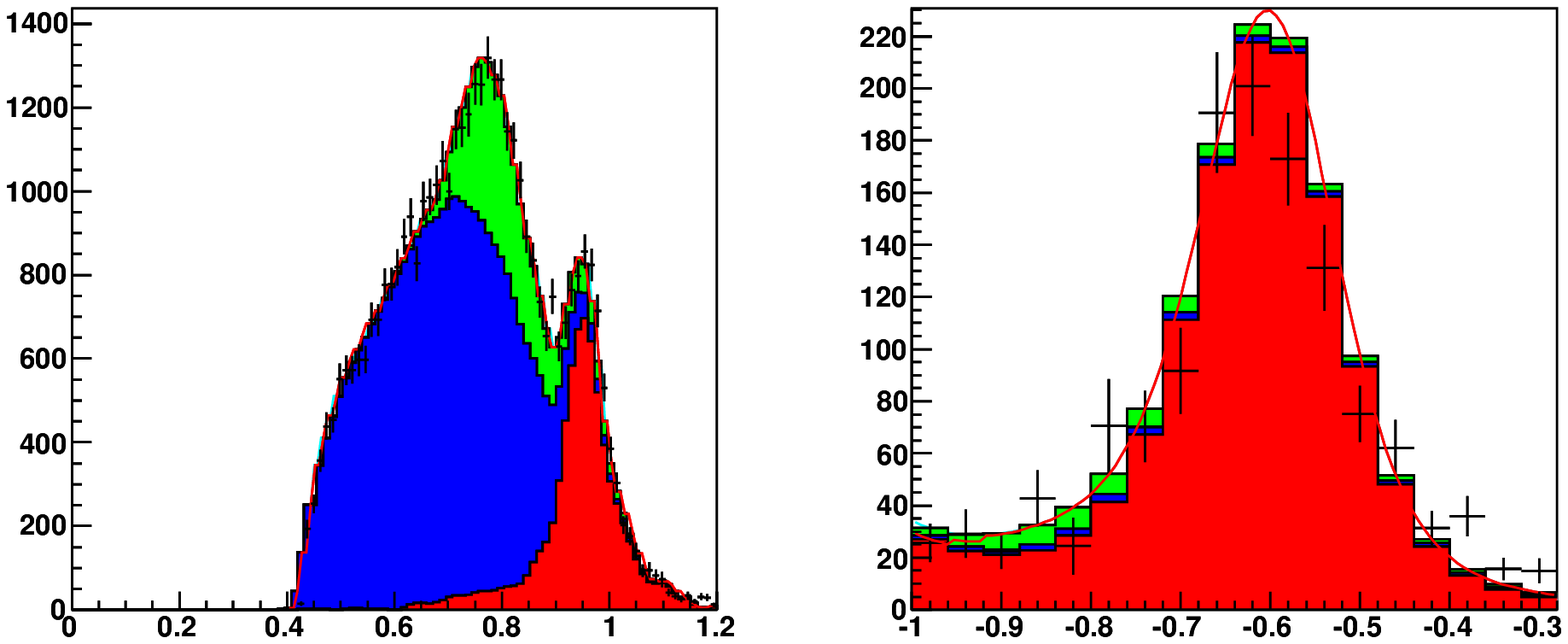}
\caption{Simultaneous fit for $m_h$=60MeV/c$^2$, $\tau_h=10^{-10} $s. Stripe 3 ($0.3 < x < 0.35$).
 $\chi^2/n.d.f. = 162.6/81$. Points with errors -- data, magenta -- signal, red -- $K_{\pi2}$, blue -- $K_{\mu3}$, green -- $K_{\mu2\gamma}$.}\label{stripe1}
\begin{picture}(1,1)
\put(-20,45){$y$}
\put(120,45){$cos\theta_{\mu\gamma}$}
\end{picture}
\end{figure}
\end{center}

\begin{center}
\begin{figure}[!h]
\centering
\includegraphics[scale=.5 , angle=0]{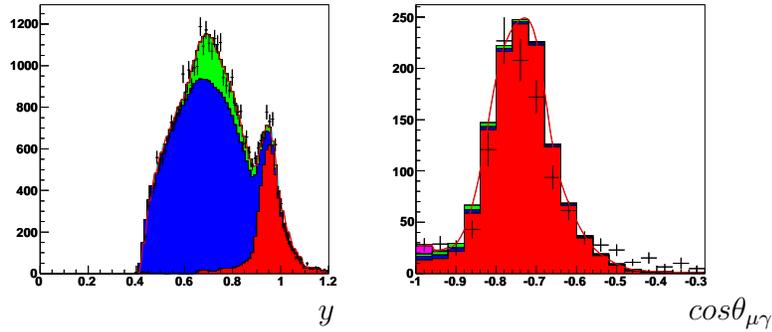}
\caption{Simultaneous fit for $m_h$=60MeV/c$^2$, $\tau_h=10^{-10} $s. Stripe 5 ($0.4 < x < 0.45$).
$\chi^2/n.d.f. = 188.3/81$. Points with errors -- data, magenta -- signal, red -- $K_{\pi2}$, blue -- $K_{\mu3}$, green -- $K_{\mu2\gamma}$.}\label{stripe5}
\begin{picture}(1,1)
\put(-20,45){$y$}
\put(120,45){$cos\theta_{\mu\gamma}$}
\end{picture}
\end{figure}
\end{center}

As an example, the results of simultaneous fits in $x$-stripes 3 and 5
($m_h$=60MeV/c$^2$, $\tau_h=10^{-10} $s, Majorana type) are shown in Fig.~\ref{stripe1},~\ref{stripe5}.
%

%

 Both signal and background shapes are taken from MC. MC histograms are smoothed
and the result is stored as $f(z)$ function ($z = y$ or $cos\theta_{\mu\gamma}$). 
For better fit, we allow these functions to be slightly widen and shifted. We do it by using $f(k\cdot z + b)$
 instead of $f(z)$ in the fit, where fit parameters $k$ and $b$
are the same for signal and background and are different for  $y$ and $cos \theta_{\mu\gamma}$. 
 For all selected $x$-stripes $k\sim1$ and $b\sim0$, i.e. our MC describes the data properly (see \cite{istra_mng} for details).

The simultaneous fit gives a signal event number in each $x$-stripe.
 As we use the same data several times we should take care of the correct estimation of a statistical error.
The whole procedure of the simultaneous fit is as follows:
\begin{itemize}
\item do simultaneous fit of two histograms and obtain $\{p_i\}$ -- best parameter values (they correspond to the global $\chi^2$ minimum);
\item take $\{p_i\}$ as initial values and perform $\chi^2/n.d.f.$ and error estimation for one histogram cos~$\theta_{\mu\gamma}$ 
using a single call of MINOS program~\cite{minuit}.
\end{itemize}

\section{Signal efficiency}

\begin{figure}[h]
\begin{minipage}[t]{0.3\textwidth}
\centering
\includegraphics[width=5cm , angle=0]{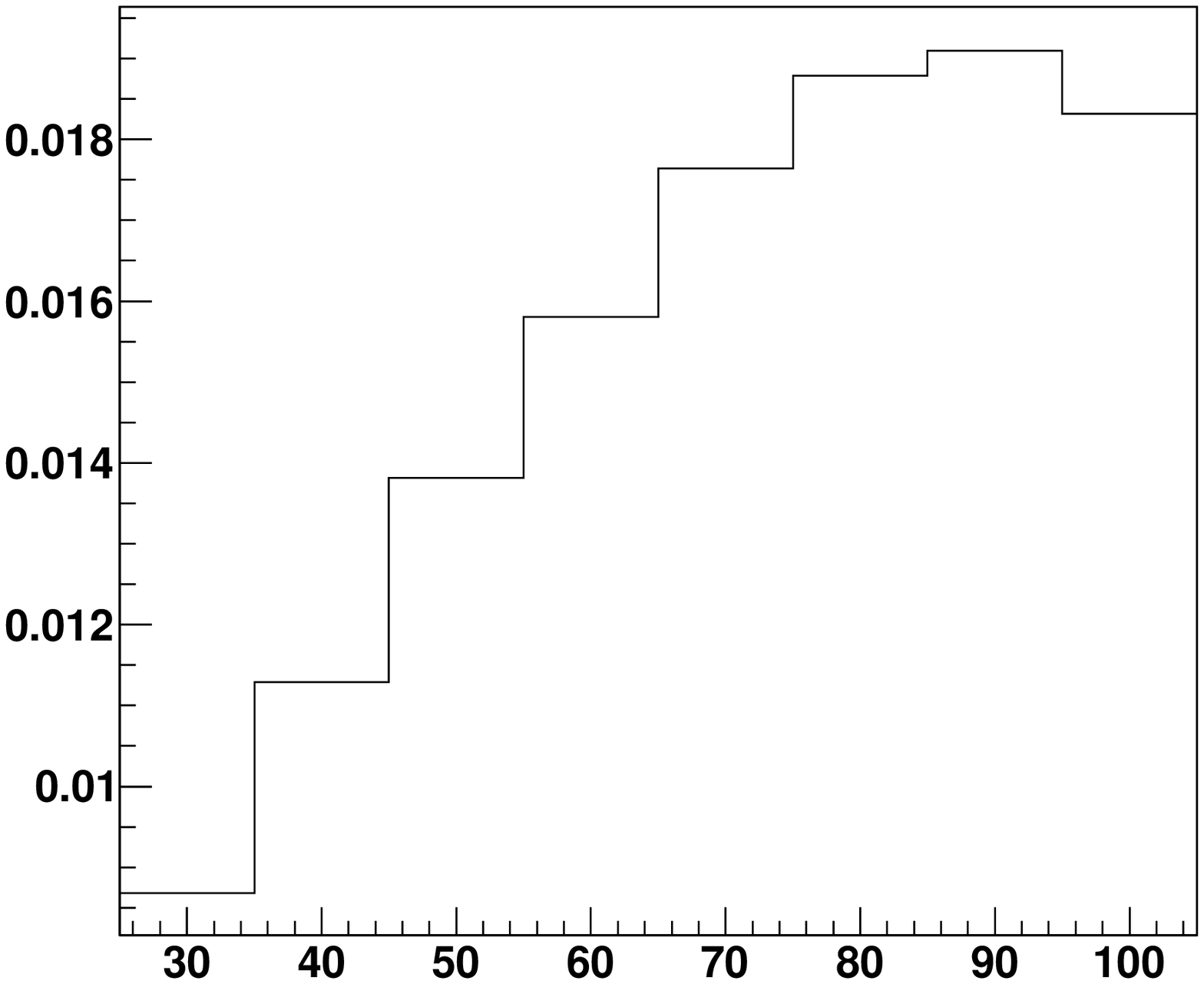}
\caption{Signal efficiency $\varepsilon$ vs $\nu_h$ mass for $\tau_h = 10^{-9}$s.}\label{eff_t09}
\end{minipage}  
\hspace{0.2cm}
\begin{minipage}[t]{0.3\textwidth}
\centering
\includegraphics[width=5cm , angle=0]{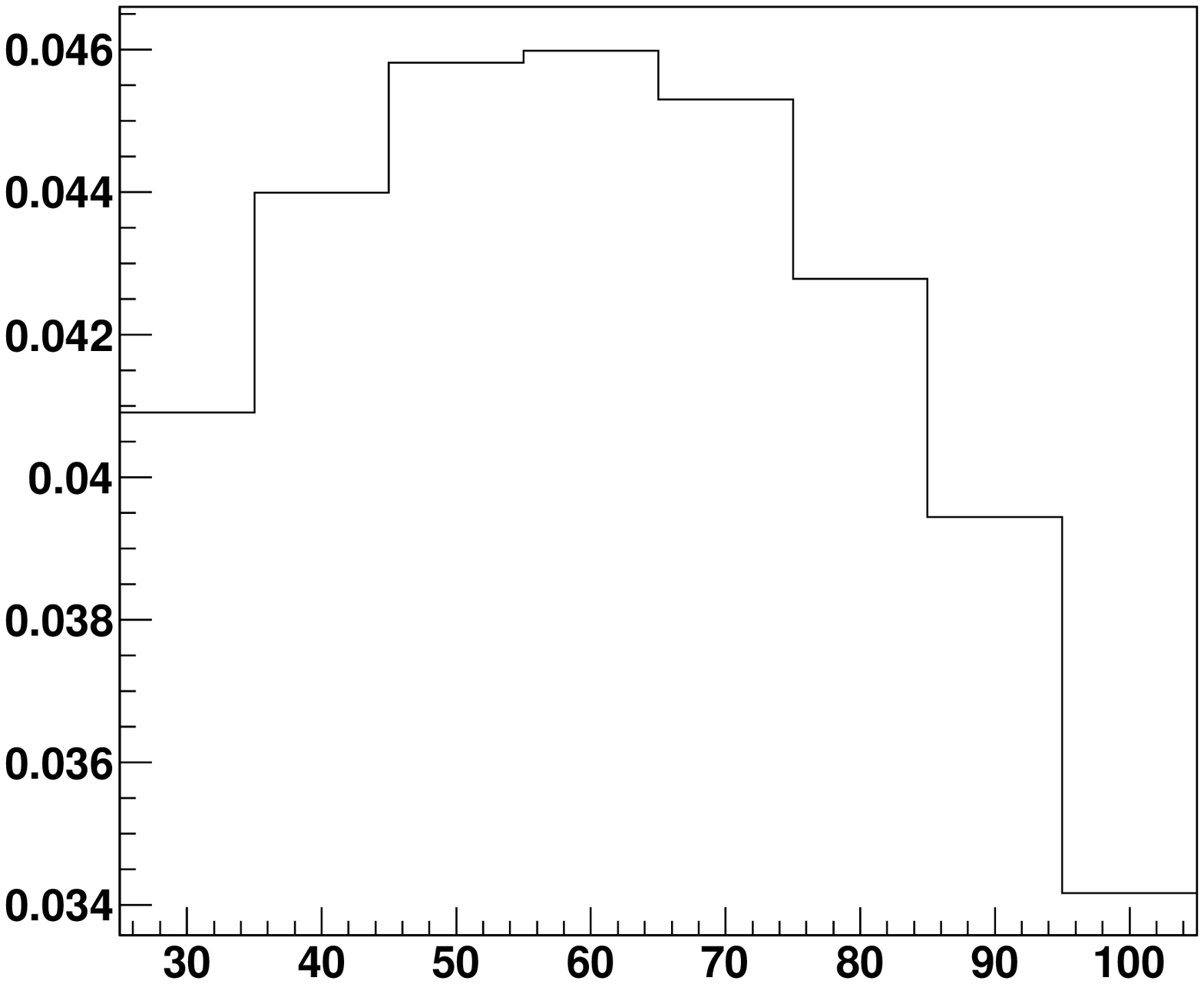}
\caption{Signal efficiency $\varepsilon$ vs $\nu_h$ mass for $\tau_h = 10^{-10}$s.}\label{eff_t10}
\end{minipage}
\hspace{0.2cm}
\begin{minipage}[t]{0.3\textwidth}
\centering
\includegraphics[width=5cm , angle=0]{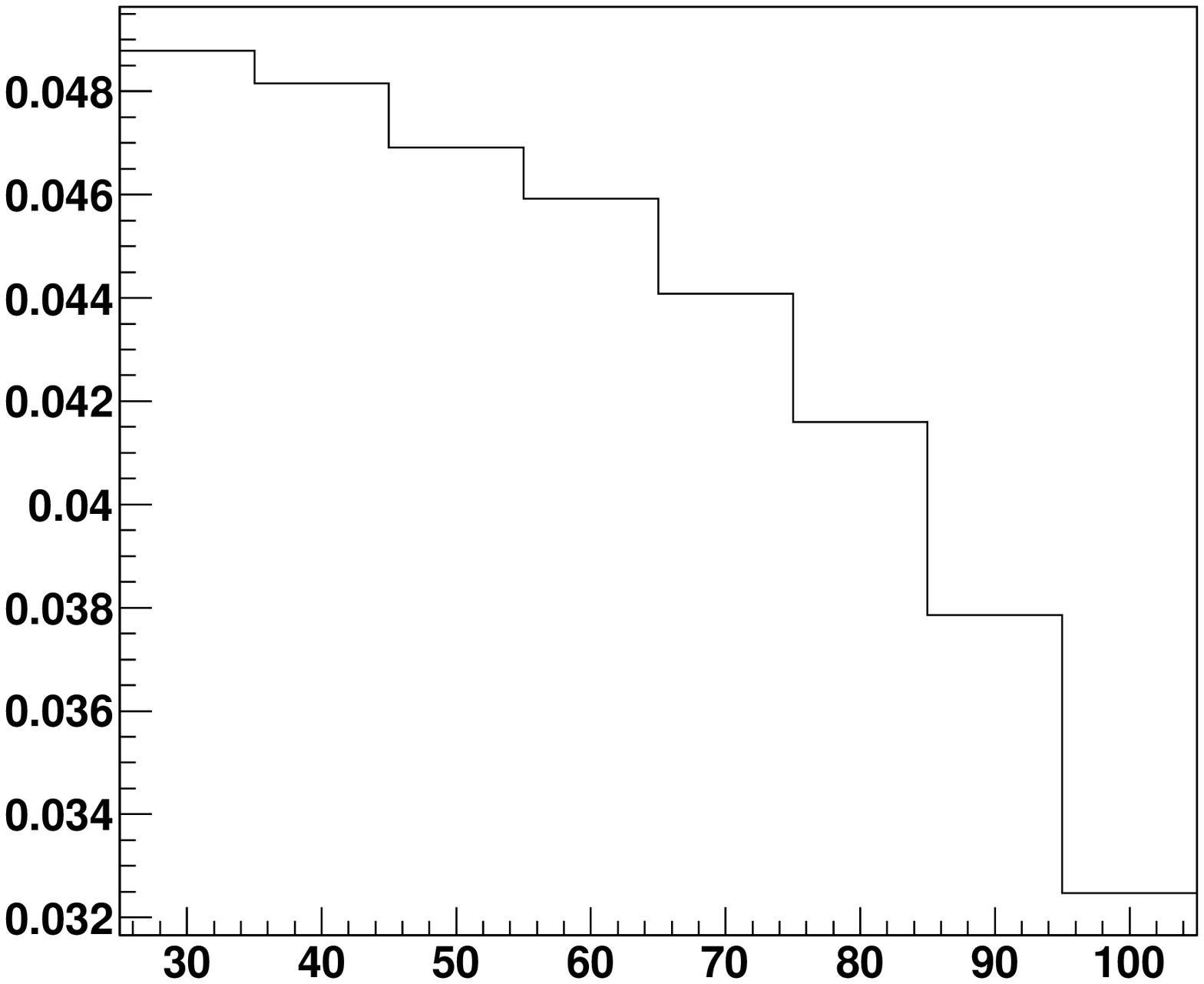}
\caption{Signal efficiency $\varepsilon$ vs $\nu_h$ mass for $\tau_h = 10^{-11}$s.}\label{eff_t11}
\end{minipage}
\begin{picture}(1,1)
\put(-30,-5){\small $m_h$, MeV/c$^2$}
\put(-180,-5){\small $m_h$, MeV/c$^2$}
\put(-330,-5){\small $m_h$, MeV/c$^2$}
\put(-137,113){\small $\varepsilon$}
\put(-287,113){\small $\varepsilon$}
\put(-437,113){\small $\varepsilon$}
\end{picture}
\end{figure}

A signal efficiency is defined as a number of events passed all cuts and fallen into the final 
kinematic region
divided by an initial event number in the same kinematic region.
Efficiency distributions for different lifetimes are shown in Figs.~\ref{eff_t09} -- \ref{eff_t11}  as a function of $\nu_h$ mass.

Two factors influence the signal efficiency:
\begin{itemize}
\item $\nu_h$ effective lifetime is larger for small $m_h$ because of a Lorentz $\gamma$-factor;
\item a signal peak in $y$ moves towards small muon energies for large $m_h$ masses and is partly cut by $y > 1$ selection; it results
in a lower efficiency for large $\nu_h$ masses.
\end{itemize}

\section{Upper limit on $|U_{\mu h}|^2$ }

\subsection{$|U_{\mu h}|^2$ extraction from $x$-stripes}

From the simultaneous fit we get $N_{exp}(K\to\mu\nu_h)$ for each $x$-stripe. This event number should be transformed into $|U_{\mu h}|^2$.
This could be done either for each $x$-stripe or for the entire selected kinematic region. To avoid a systematics depending on an $x$-stripe
(for example backgrounds could be described worse by MC in a certain $x$-stripe and shift a total result) we use the first method.

As a normalization it is natural to use $K_{\mu2\gamma}$ decay. The main formula is the following: 
\begin{center}
$\myfrac{N_{exp}(K\to\mu\nu_h)}{N_{exp}(K_{\mu2\gamma})} = \myfrac{BR(K\to\mu\nu_h)}{BR(K_{\mu2\gamma})} \myfrac{\varepsilon(K\to\mu\nu_h)}{\varepsilon(K_{\mu2\gamma})} $. 
\end{center}

$N_{exp}(K_{\mu2\gamma})$ is taken from our previous analysis \cite{istra_mng} in a wide kinematic region,
 $\varepsilon(K\to\mu\nu_h)$ and $\varepsilon(K_{\mu2\gamma})$
are efficiencies obtained from MC. $BR(K_{\mu2\gamma})$ is taken from the theory because an experimental measurement is very old and has
a large error (the mean value is consistent with the theoretical prediction). In future, this $BR$ could be measured using ISTRA+ data.    

$BR(K\to\mu\nu_h)$ is substituted by the following expression: 
\begin{center}
$BR(K\to\mu\nu_h) = BR(K_{\mu2}) \cdot |U_{\mu h}|^2 \cdot f(m_h) $.
\end{center}
Here $BR(K_{\mu2})$ is taken from PDG \cite{pdg}, $f(m_h)$ contains chirality flip and phase space factors. 
For the Dirac case from the general
formula in \cite{gorby} we get:
\begin{center}
$f_D(m_h) = \myfrac{m_h^2 (1 - \myfrac{m_h^2}{m_K^2} + 2\myfrac{m_\mu^2}{m_K^2} + \myfrac{m_\mu^2}{m_h^2}(1-\myfrac{m_\mu^2}{m_K^2}))}{m_\mu^2(1-\myfrac{m_\mu^2}{m_K^2})^2} \cdot \sqrt{(1 + \myfrac{m_h^2}{m_K^2}-\myfrac{m_\mu^2}{m_K^2})^2 - 4\myfrac{m_h^2}{m_K^2}}$.
\end{center}

%
\begin{center}
\begin{figure}[h]
\centering
\includegraphics[scale=.3 , angle=0]{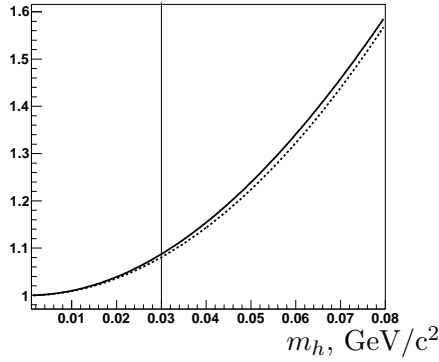}
\caption{Factor $f_D(m_h)$ (solid line) and $(1+\myfrac{m_h^2}{m_\mu^2})$(dashed).}\label{factor}
\begin{picture}(1,1)
\put(30,40){\small $m_h$, GeV/c$^2$}
\end{picture}
\end{figure}
\end{center}

 $f_D(m_h)$  is dominated by the chirality flip factor
 $1+(\myfrac{m_h}{m_\mu})^2$  (see Fig.~\ref{factor}) and for the mass interval $m_h$ = 30-80MeV/c$^2$ it varies from 1.1 to 1.6.
 For the Majorana case, $f_M(m_h) = 2 \cdot f_D(m_h)$.

Finally for $|U_{\mu h}|^2$  we get 
\begin{center}
$|U_{\mu h}|^2  = \myfrac{N_{exp}(K\to\mu\nu_h)}{N_{exp}(K_{\mu2\gamma})}
\myfrac{BR(K_{\mu2\gamma})}{BR(K_{\mu2})} 
\myfrac{\varepsilon(K_{\mu2\gamma})}{\varepsilon(K\to\mu\nu_h)}
\myfrac{1}{f(m_h)}$.
\end{center}

\begin{center}
\begin{figure}[h]
\begin{minipage}[t]{0.5\textwidth}
\centering
\includegraphics[scale=.3 , angle=0]{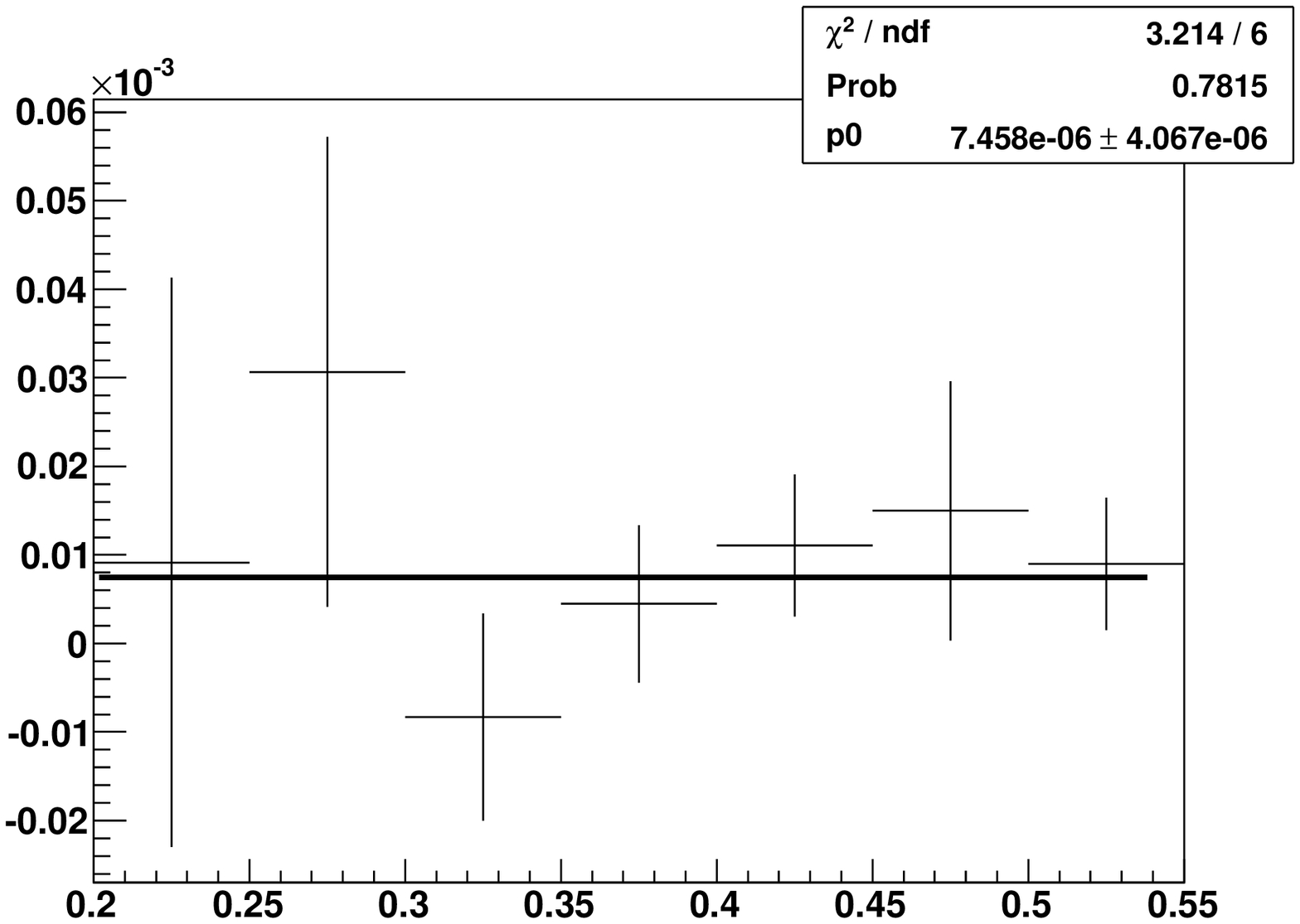}
\caption{Final fit, $m_h$=50MeV/c$^2$, $\tau_h=10^{-10}$s. Dirac case.}\label{average2}
\end{minipage}  
\hspace{0.2cm}
\begin{minipage}[t]{0.5\textwidth}
\centering
\includegraphics[scale=.3 , angle=0]{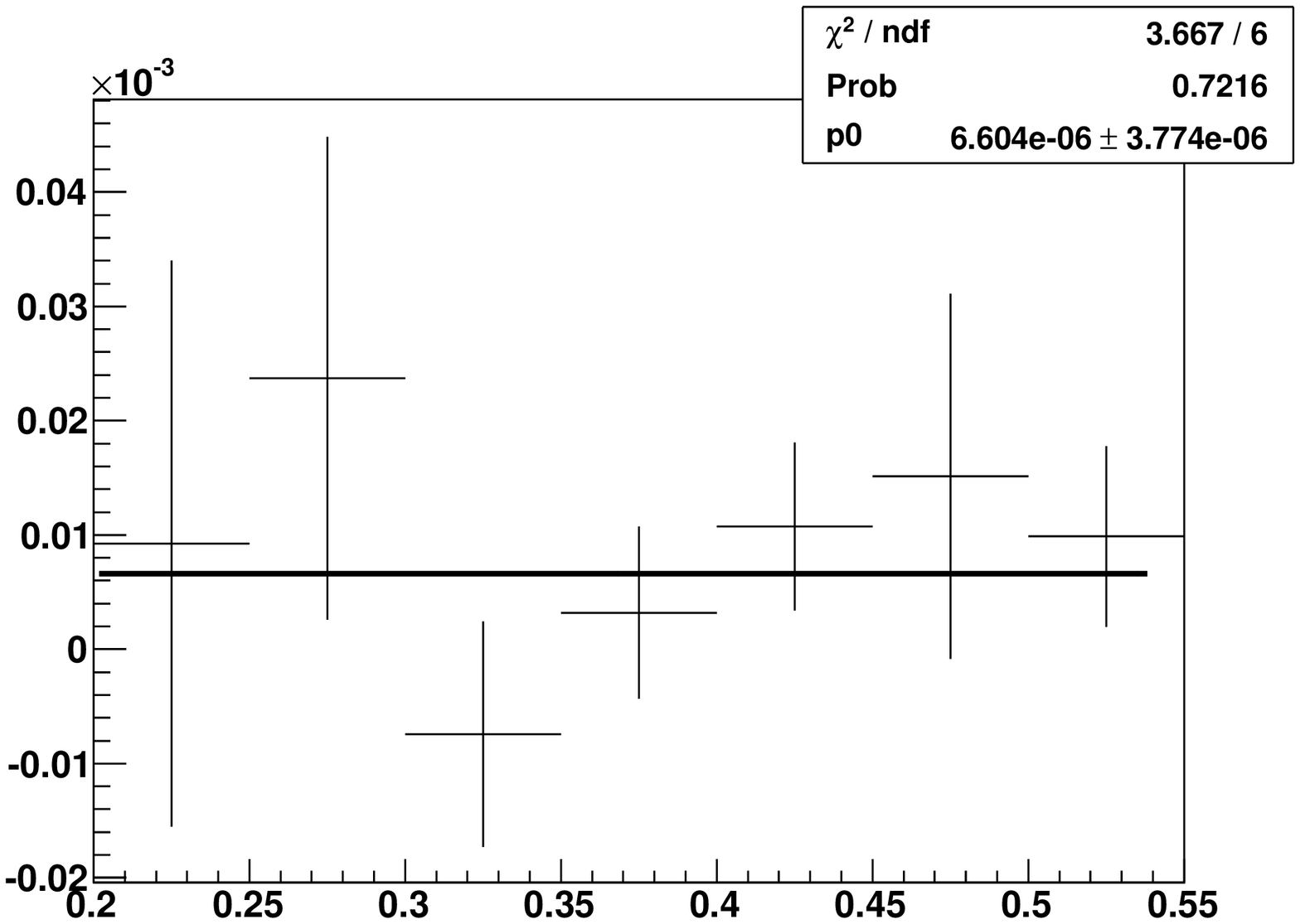}
\caption{Final fit, $m_h$=50MeV/c$^2$, $\tau_h=10^{-10}$s. ~~~Majorana case.}\label{average}
\end{minipage}  
\begin{picture}(1,1)
\put(165,35){\small $x$}
\put(0,135){\small $|U^{str}_{\mu h}|^2$}
\put(405,35){\small $x$}
\put(240,135){\small $|U^{str}_{\mu h}|^2$}
\end{picture}
\end{figure}
\end{center}

\subsection{Averaging $|U_{\mu h}|^2$ and setting upper limits}

Values $|U_{\mu h}|^2$ are calculated for all seven $x$-stripes (we will denote them as $|U^{str}_{\mu h}|^2$) 
and then averaged ($|U^{av}_{\mu h}|^2$). We call this averaging procedure
a final fit.
The final fit for a certain $(m_h, \tau_h)$ is shown in Fig.~\ref{average2} (Dirac case) and Fig.~\ref{average}
(Majorana case).
The effect in $|U^{av}_{\mu h}|^2$ (only a statistical error is considered here) does not exceed $(1-2)\sigma$
for all $(m_h, \tau_h)$ and hence an upper limit should be set.
The averaged value is used for a one-sided upper limit (U.L.) calculation: 
\begin{center}
$U.L.(95\% C.L.) = |U^{av}_{\mu h}|^2 + 1.64 \cdot \sigma_{U,tot} $.
\end{center}
Here $\sigma_{U,tot}$ is a total error of $|U^{av}_{\mu h}|^2$ measurement.

\subsection{Systematic error}

Main sources of the systematics are:
\begin{itemize}
\item fit systematics;
\item cut on $x$ (number of $x$-stripes in the final fit);
\item cut on $y$ in $x$-stripes;
\item $x$-stripe width;
\item bin size in $y$ and $cos\theta_{\mu\gamma}$ histograms;
\item cut on $z_{vtx}$.
\end{itemize}

The largest contribution to the total systematic error comes from the {\it fit systematics} caused by 
non-ideal MC shapes of the signal and the backgrounds. To estimate this systematic error the following procedure is used:
\begin{itemize}
\item errors for $|U^{str}_{\mu h}|^2$ are scaled for each $x$-stripe proportionally to $\sqrt{\chi^2/n.d.f.}$;
\item averaging is repeated with these new scaled errors;
\item new averaged value $|U^{av,scaled}_{\mu h}|^2$ has larger error $\sigma_{scaled}$ which is treated as 
$\sigma_{scaled} = \sqrt{\sigma_{stat}^2 + \sigma_{syst, fit}^2}$. Here $\sigma_{stat}$ is a statistical error
of  $|U^{av}_{\mu h}|^2$.
\end{itemize}

The systematics of a {\it cut on $x$} is estimated as follows: 
\begin{itemize}
\item averaging is done for different number of $x$-stripes in the fit (varying cut on $x$);
\item the dependence of $|U^{av}_{\mu h}|^2$ on $x$-cut is fitted by a straight line;
\item the slope of this line multiplied by the $x$-stripe width is the estimation of the systematic error.
\end{itemize}
Details of this procedure are described in \cite{istra_mng}.

The systematics of a {\it cut on $y$} is calculated in a similar way by varying 
$y$-cut value and fitting the dependence of $|U^{av}_{\mu h}|^2$ on $y$-cut by a straight line.

{\it $ x$-stripe width} is changed ($dx = 0.035, dx = 0.07$) and the whole procedure (simultaneous fits
in $x$-stripes, final fit) is repeated for new $dx$. New values of $|U^{av,new}_{\mu h}|^2$ are compatible
with old ones and hence no systematics is found here.

The systematics caused by the {\it bin size in $y$ and $cos~\theta_{\bf \mu\gamma}$ histograms} 
is estimated in a similar way and the result is the same: no additional error is found.

The {\it cut on $z_{vtx}$} is varied within the resolution. The systematical error is calculated similar to that of the cut on $x$ .

\subsection{Upper limits}

\begin{figure}[!h]
\begin{minipage}[t]{0.3\textwidth}
\centering
\includegraphics[width=5cm , angle=0]{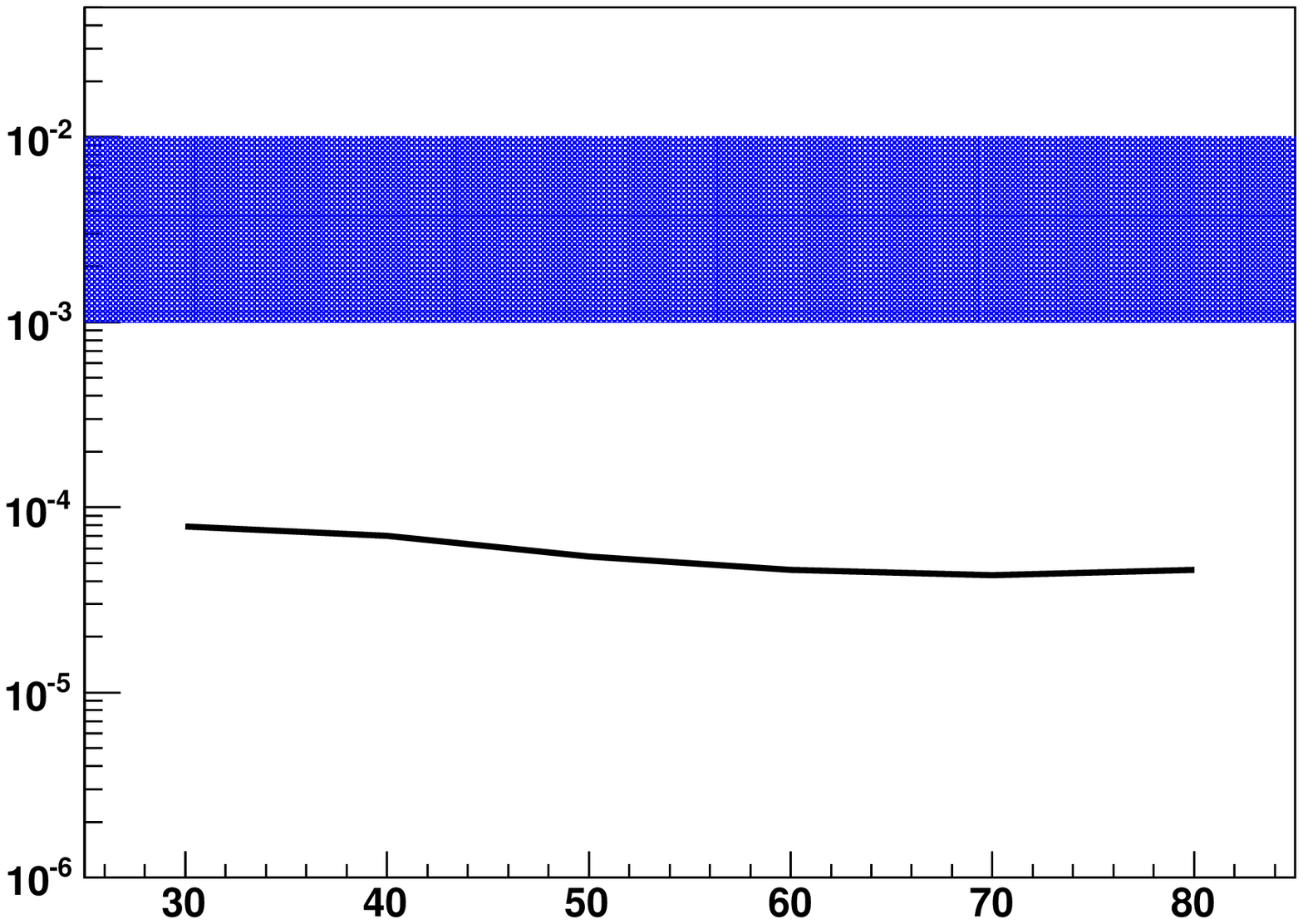}
\end{minipage} 
\hspace{0.2cm}
\begin{minipage}[t]{0.3\textwidth}
\centering
\includegraphics[width=5cm , angle=0]{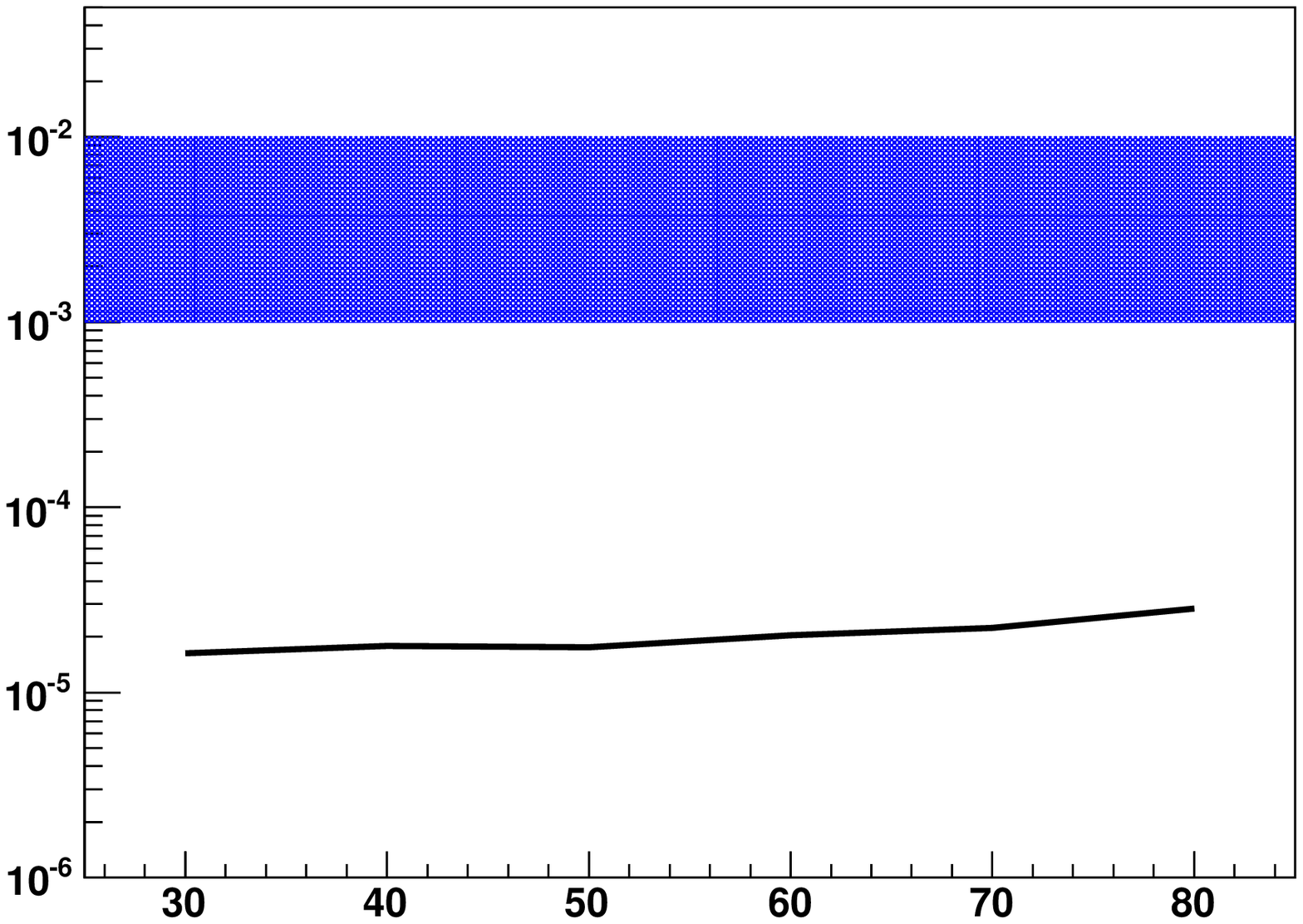}
\end{minipage}
\hspace{0.2cm}
\begin{minipage}[t]{0.3\textwidth}
\centering
\includegraphics[width=5cm , angle=0]{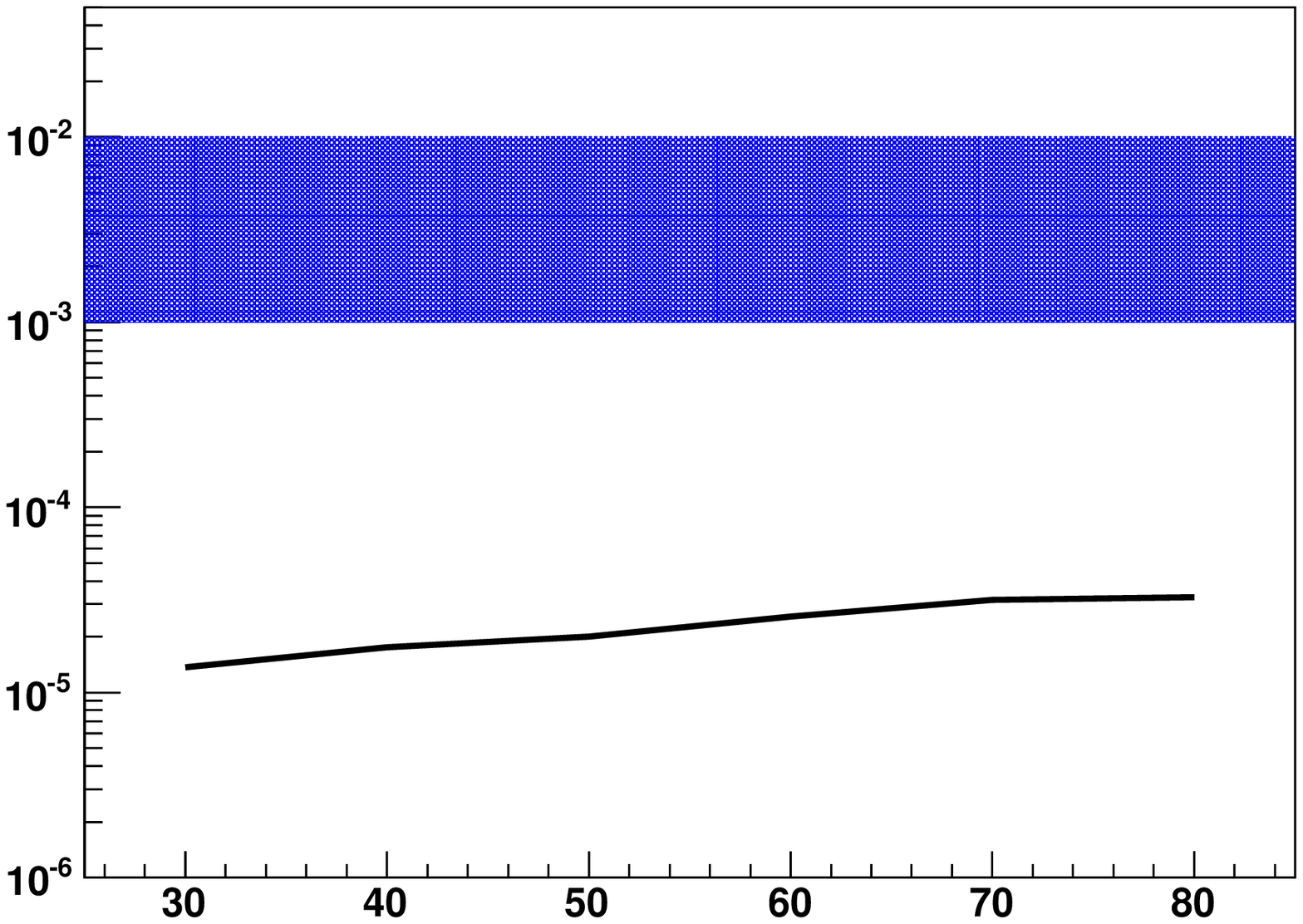}
\end{minipage} 
\caption{Upper limit for $|U_{\mu h}|^2$ vs $m_h$ (Dirac case). Black line -- obtained upper limits, blue stripe -- prediction from \cite{gninenko}.}\label{ul_dirac}
\begin{picture}(1,1)
\put(105,47){\footnotesize $m_h$, MeV/c$^2$}
\put(255,47){\footnotesize $m_h$, MeV/c$^2$}
\put(402,47){\footnotesize $m_h$, MeV/c$^2$}
\put(45,142){\footnotesize $\tau_h = 10^{-9}$s}
\put(195,142){\footnotesize $\tau_h = 10^{-10}$s}
\put(345,142){\footnotesize $\tau_h = 10^{-11}$s}
\put(-10,130){\small $U.L.$}
\put(140,130){\small $U.L.$}
\put(290,130){\small $U.L.$}
\end{picture}
\end{figure}

\begin{figure}[!h]
\begin{minipage}[t]{0.3\textwidth}
\centering
\includegraphics[width=5cm , angle=0]{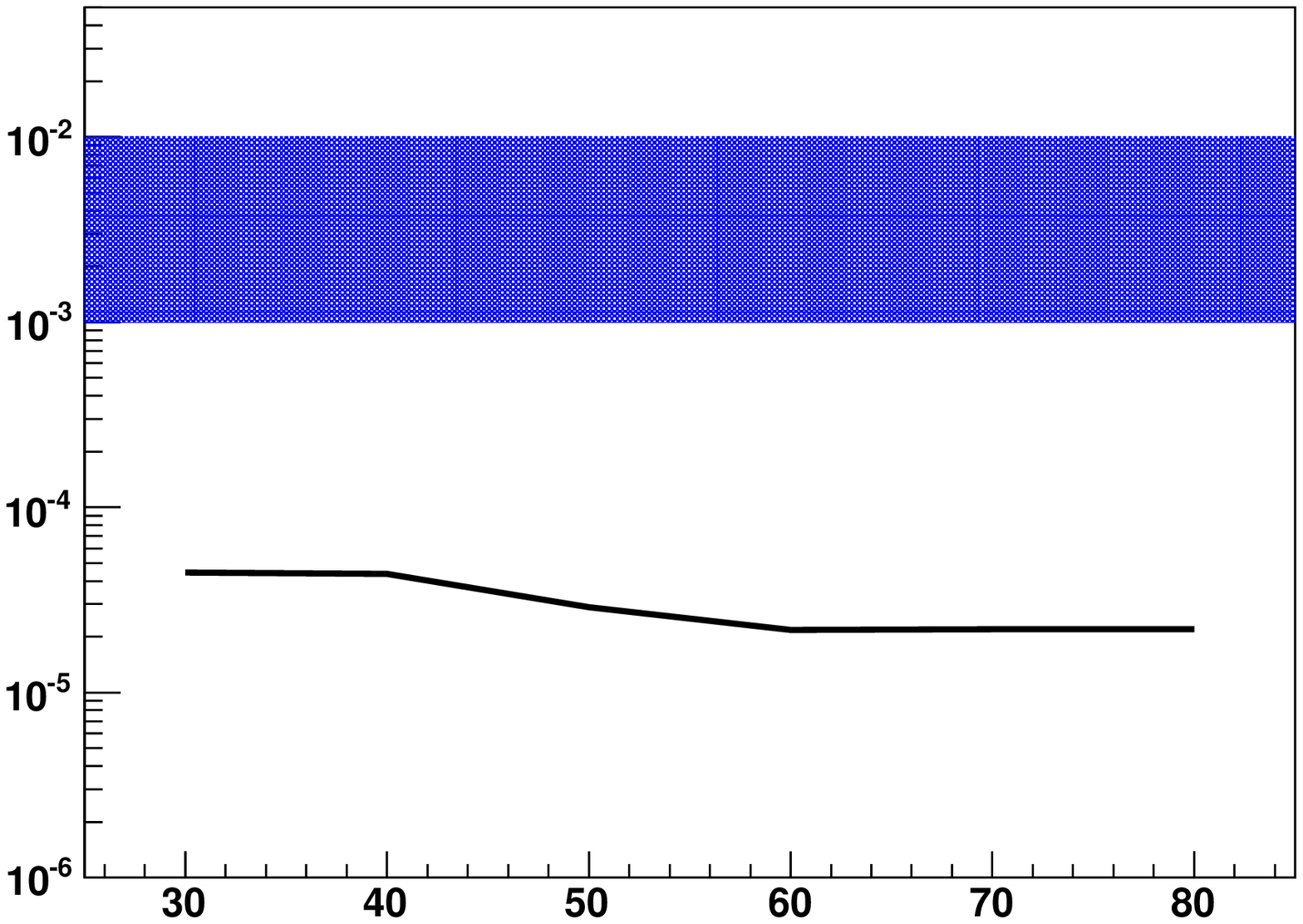}
\end{minipage} 
\hspace{0.2cm}
\begin{minipage}[t]{0.3\textwidth}
\centering
\includegraphics[width=5cm , angle=0]{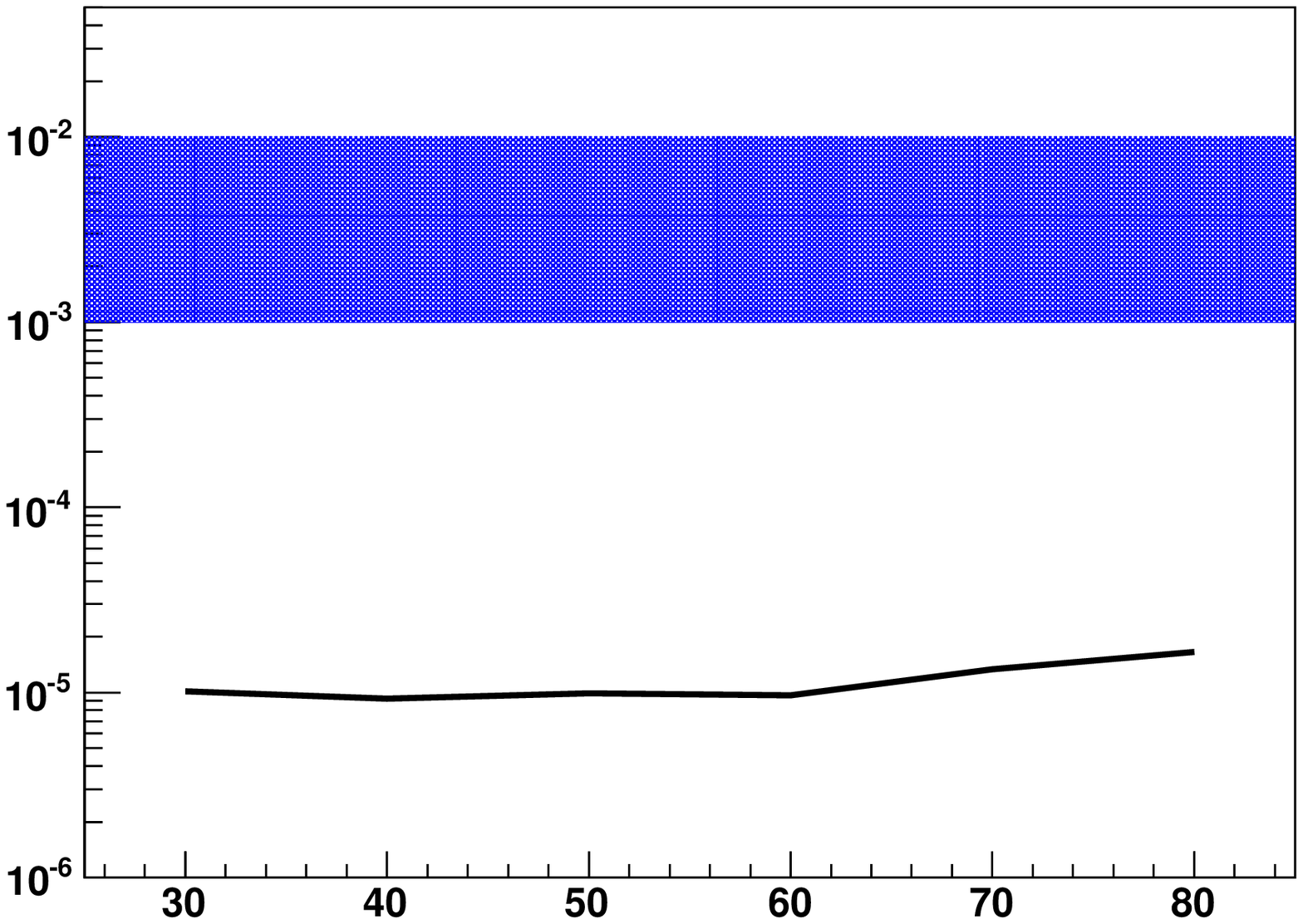}
\end{minipage}
\hspace{0.2cm}
\begin{minipage}[t]{0.3\textwidth}
\centering
\includegraphics[width=5cm , angle=0]{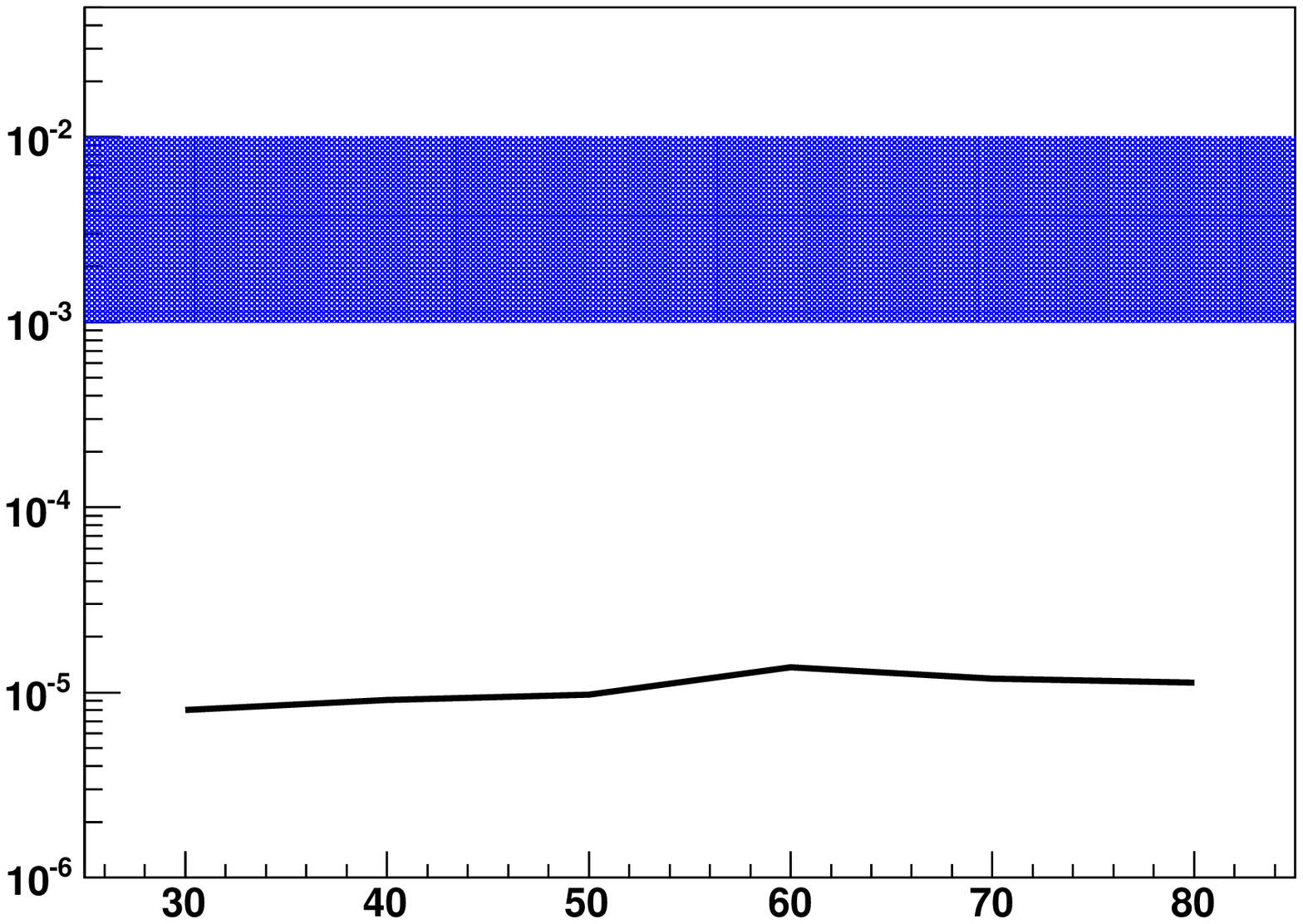}
\end{minipage} 
\caption{Upper limit for $|U_{\mu h}|^2$ vs $m_h$ (Majorana case). Black line -- obtained upper limits, blue stripe -- prediction from \cite{gninenko}.}\label{ul_maj}
\begin{picture}(1,1)
\put(105,47){\footnotesize $m_h$, MeV/c$^2$}
\put(255,47){\footnotesize $m_h$, MeV/c$^2$}
\put(402,47){\footnotesize $m_h$, MeV/c$^2$}
\put(45,142){\footnotesize $\tau_h = 10^{-9}$s}
\put(195,142){\footnotesize $\tau_h = 10^{-10}$s}
\put(345,142){\footnotesize $\tau_h = 10^{-11}$s}
\put(-10,130){\small $U.L.$}
\put(140,130){\small $U.L.$}
\put(290,130){\small $U.L.$}
\end{picture}
\end{figure}

Upper limits for different lifetimes as a function of $m_h$ are shown in Figs.~\ref{ul_dirac}, \ref{ul_maj}.
The limits are calculated for the following values of  $m_h$: 30, 40, 50, 60, 70 and 80 MeV/c$^2$. The curve in the figures is the interpolation
between these values.
The upper limits could be compared with the region predicted in \cite{gninenko} (shown with a blue stripe).
As an example, contribution of all errors to the final result is shown in Table \ref{syst_dirac} for the Dirac case, $\tau_h = 10^{-10}$s.

\vspace{0.2cm}

\vspace{0.5cm} 

Exact numbers are collected in Tables~\ref{results_dirac} and \ref{results_majorana}.
\vspace{0.5cm}

\begin{table}[!h]\footnotesize
\centering
    \begin{tabular}{|l|l|l|l|l|l|l|l|}  
      \hline  
   
        $m_h$, MeV/c$^2$  & $|U_{\mu h}|^2$ & $\sigma_{stat}$  &  $\sigma_{syst,x}$ &  $\sigma_{syst,y}$ &  $\sigma_{syst,z}$&   $\sigma_{syst,fit}$   &   $U.L.$ \\
         \hline  
          
30  & 0.6   & 0.4  &  0.001   & 0.002 & 0.1    & 0.5    & 1.6   \\ 
40  & 0.7   & 0.4  &  0.01   & 0.009  & 0.04   & 0.5    & 1.8   \\ 
50  & 0.7   & 0.4  &  0.01   & 0.004  & 0.06   &  0.5   &  1.8   \\ 
60  & 0.8   & 0.4  & 0.005    & 0.03  & 0.4    & 0.5    & 2.0   \\ 
70  & 0.9   & 0.5  & 0.0002   & 0.2   & 0.4    & 0.5    & 2.2   \\ 
80  & 1.0   & 0.5  &  0.08   & 0.3    & 0.7    &    0.6  &   2.8  \\ 
                    
	  \hline

       \end{tabular}  
	\caption{Statistical and systematic errors for $\nu_h$ of the Dirac type, $\tau_h = 10^{-10} $s. Fit results and $U.L.$'s are in $10^{-5}$ units.}\label{syst_dirac}
\end{table}

\begin{table}[!h]\footnotesize
\centering
    \begin{tabular}{|l|l|l|l|}  
      \hline  
   
        $m_h$, MeV/c$^2$  & $U.L., \tau_h = 10^{-9} $s & $U.L., \tau_h = 10^{-10} $s  &  $U.L., \tau_h = 10^{-11} $s    \\
         \hline  
          
30  & 7.9   & 1.6  &  1.4     \\ 
40  & 7.0   & 1.8  &  1.8     \\ 
50  & 5.4   & 1.8  &  2.0     \\ 
60  & 4.6   & 2.0  &  2.6     \\ 
70  & 4.3   & 2.2  &  3.2    \\ 
80  & 4.6   & 2.8  &  3.3    \\ 
                    
	  \hline

       \end{tabular}  
	\caption{Upper limits for $\nu_h$ of the Dirac type. $U.L.$'s are in $10^{-5}$ units.}\label{results_dirac}
\end{table}

\begin{table}[!h]\footnotesize
\centering
    \begin{tabular}{|l|l|l|l|}  
      \hline  
   
        $m_h$, MeV/c$^2$  & $U.L., \tau_h = 10^{-9} $s & $U.L., \tau_h = 10^{-10} $s  &  $U.L., \tau_h = 10^{-11} $s    \\
         \hline  
          
30  & 4.5   & 1.0  &  0.8     \\ 
40  & 4.4   & 0.9  &  0.9     \\ 
50  & 2.9   & 1.0  &  1.0     \\ 
60  & 2.2   & 1.0  &  1.4     \\ 
70  & 2.2   & 1.3  &  1.2    \\ 
80  & 2.2   & 1.7  &  1.1    \\ 
                    
	  \hline

       \end{tabular}  
	\caption{Upper limits for $\nu_h$ of the Majorana type. $U.L.$'s are in $10^{-5}$ units.}\label{results_majorana}
\end{table}

\section{Conclusions}
We have performed a search for a heavy neutrino of the Dirac and Majorana type in $K\to\mu\nu_h (\nu_h\to\nu\gamma)$ decay 
assuming that $\nu_h$ is a part of $\nu_\mu$
 flavor eigenstate and decays radiatively into a massless neutrino and a photon
and obtained upper limits at $95\%$ C.L. for
the mixing matrix element $|U_{\mu h}|^2$. 

The upper limit at 95$\%$ C.L. in a mass region $30MeV/c^2 \le m_h \le$ 80MeV/c$^2$ for $10^{-11}s \le \tau_h \le 10^{-9}$s is  $U.L. \sim (1\div5) \cdot 10^{-5}$
(Majorana type of $\nu_h$) and  $U.L. \sim (2\div8) \cdot 10^{-5}$ (Dirac type).
The obtained values close the allowed region for $|U_{\mu h}|^2$ suitable for LSND/KARMEN/MiniBooNE anomaly explanation proposed in \cite{gninenko}.

Authors would like to thank S.N.Gninenko, D.S.Gorbunov, V.A.Rubakov, A.A.Saratov (INR RAS) and R.Shrock (YITP, Stony Brook) for numerous discussions.
 The work is supported by the Russian Fund for Basic Research (grants 10-02-00330-a and 11-02-00870-a).

\end{document}